\documentclass{aa}  

\usepackage{natbib}
\usepackage{graphicx}
\usepackage{txfonts}
\usepackage{hyperref}

\usepackage{color}

\usepackage{upgreek} \newcommand\micron{\ensuremath{\rm \upmu m}\@\xspace} 

\begin{document}

\title{Atmospheres and wind properties of non-spherical AGB stars}

\author{S. Liljegren\inst{1}
\and S. H{\"o}fner \inst{1}
\and B. Freytag\inst{1}
\and S. Bladh\inst{1}
}

\institute{$^1$Division of Astronomy and Space Physics, Department of Physics and Astronomy, Uppsala University, Box 516, 751 20 Uppsala, Sweden\\
{\email{sofie.liljegren@physics.uu.se}}\\          
           }
\date{}

\abstract{
    The wind-driving mechanism of asymptotic giant branch (AGB) stars is commonly attributed to a two-step process: first, gas in the stellar atmosphere is levitated by shockwaves caused by stellar pulsation, then accelerated outwards by radiative pressure on newly formed dust, inducing a wind. Dynamical modelling of such winds usually assumes a spherically symmetric star. 
}{
    We explore the potential consequences of complex stellar surface structures, as predicted by three-dimensional (3D) star-in-a-box modelling of M-type AGB stars, on the resulting wind properties with the aim to improve the current wind models. 
}{
    Two different modelling approaches are used; the CO$^5$BOLD 3D star-in-a-box code to simulate the convective, pulsating interior and lower atmosphere of the star, and the DARWIN one-dimensional (1D) code to describe the dynamical atmosphere where the wind is accelerated. The gas dynamics of the inner atmosphere region at distances of $R\sim1-2R_\star$, which both modelling approaches simulate, are compared. Dynamical properties and luminosity variations derived from CO$^5$BOLD interior models are used as input for the inner boundary in DARWIN wind models in order to emulate the effects of giant convection cells and pulsation, and explore their influence on the dynamical properties. 
}{
    The CO$^5$BOLD models are inherently anisotropic, with non-uniform shock fronts and varying luminosity amplitudes, in contrast to the spherically symmetrical DARWIN wind models. DARWIN wind models with CO$^5$BOLD-derived inner boundary conditions produced wind velocities and mass-loss rates comparable to the standard DARWIN models, however the winds show large density variations on time-scales of 10-20 years. 
    }
    {The method outlined in this paper derives pulsation properties from the 3D star-in-a-box CO$^5$BOLD models, to be used in the DARWIN models. If the current grid of CO$^5$BOLD models is extended, it will be possible to construct extensive DARWIN grids with inner boundary conditions derived from 3D interior modelling of convection and pulsation, and avoid the free parameters of the current approach.}


   \keywords{Stars: AGB and post-AGB - Stars: atmospheres - Stars: winds, outflows - Stars: oscillations (including pulsations) - shockwaves}

\maketitle

\section{Introduction}

Low- to intermediate-mass stars, that is, 1-8 $M_\odot$ on the zero age main sequence (ZAMS), will lose a large portion of their mass during the asymptotic giant branch (AGB) phase, before turning into white dwarfs. 
The mass loss on the AGB is due to a slow atmospheric stellar wind, driven by radiation pressure on dust in the stellar atmosphere. 
Outward-propagating shockwaves caused by the pulsation of the AGB star will periodically levitate gas to distances where the conditions are favourable for dust formation, that is, with low temperature and high density. 
The dust particles will be accelerated outwards by interacting with the radiation field, through absorption and scattering, depending the type of dust. 
Momentum is transferred from the dust to the gas through collisions, inducing a wind. 
Observationally, this scenario is supported by, for example, high-resolution spectroscopy \citep[see][]{hinkle_time_1982,scholz_derivation_2000,nowotny_line_2010} and by high-angular interferometry modelling, and imaging (see e.g. for O-rich stars: \citealt{chandler_asymmetries_2007,karovicova_new_2013,ohnaka_spatially_2012,ohnaka_clumpy_2016,ohnaka_clumpy_2017}; while for C-rich stars: \citealt{ohnaka_temporal_2007,ohnaka_asymmetric_2008,ohnaka_amber-naco_2015,sacuto_observing_2011,rau_modelling_2015,rau_adventure_2017,wittkowski_aperture_2017}).

Dynamical atmosphere models of AGB stars are used to simulate the wind-driving mechanism \citep[for a review see][]{hofner_mass_2018}. 
These wind models typically have an inner boundary situated just below the photosphere of the star, and reach out to around $20-30R_\star$, incorporating both the regions of dust formation and wind acceleration. 
Such models usually do not include any description of the stellar interior, where the pulsations originate.
Therefore, the variations of the stellar surface layers and of the luminosity, which play a key role for the wind mechanism, are typically described by a parameterised inner boundary condition.
Such an inner boundary condition was previously commonly assumed to be sinusoidal variations of both the stellar radius and the luminosity \citep[e.g.][]{hofner_dynamic_2003,hofner_dynamic_2016-1}.  
However both one-dimensional (1D) pulsation models \citep[e.g.][]{ireland_dynamical_2011} and observations \citep[e.g.][]{nowotny_line_2010,lebzelter_shapes_2011} suggest that this approach is an oversimplification. 
Previous studies have shown that assumptions made about the inner boundary may have substantial effects on both the dynamical properties of the resulting models and on the derived observables \citep{liljegren_dust-driven_2016,liljegren_pulsation-induced_2017}.

A more realistic approach to predicting the mass-loss rate and wind velocity should ideally describe the dynamics of the stellar surface and the luminosity without free parameters, and be derived from variations of pulsation models. 
The stellar interior where these variations originate, an optical thick region dominated by convection, has however proven difficult to model. 
Historically the stellar envelope region has been modelled using 1D self-excited pulsation models, with mixing length theory for describing the convective motions. 
The mixing length description is often quoted as a shortcoming of such an approach \citep[for a detailed discussion see][]{barthes_pulsation_1998}. Recent findings further suggest that 1D radial pulsation models reproduce periods well for early AGB stars, which pulsate in overtone modes but are unreliable for evolved AGB stars, and Miras, which are thought to be fundamental pulsators \citep{trabucchi_new_2017-1}.

A different approach to modelling the pulsation process of the more evolved AGB stars was explored by \cite{freytag_three-dimensional_2008} and \cite{freytag_global_2017}, with 3D star-in-a-box models of AGB stars simulated using the CO$^5$BOLD (COnservative COde for the COmputation of COmpressible COnvection in a BOx
of L Dimensions, L=2,3) code.

With a realistic 3D hydrodynamical description, the turbulent convective flows are modelled directly, avoiding crude recipes such as the mixing length theory.  
Pulsations emerge in the 3D models, with realistic periods for Mira stars \citep{freytag_global_2017}.

The CO$^5$BOLD  models encompass part of the atmosphere, with the outer boundary situated at $\sim 2 R_\star$. 
The AGB models presented by \cite{freytag_global_2017} do not include dust formation and therefore no wind driving. 
While there are plans to expand the current CO5BOLD modelling setup to include dust formation and the wind-driving region, such models will be very time consuming (the runtime is already typically a couple of CPU years per model, for parallelised code). 
They will therefore be impractical for extensive wind model grids used to derive wind properties for a wide range of stellar parameters. 

In this paper, we instead aim at improving the current 1D atmosphere and wind models (DARWIN code) by quantifying the dynamical behaviour in existing 3D convection and pulsation models (CO$^5$BOLD code) and applying the results in the 1D models.
This is done by comparing the lower atmosphere for both modelling methods, which is a region in the range $\sim1-2R_\star$ where shockwaves induced by stellar pulsation dominate the dynamics before dust has condensed. 
The impact that the non-spherical morphology seen in the  CO$^5$BOLD might have on the wind properties is discussed. 
The 3D models are then used to derive new inner boundary conditions for 1D atmosphere models, with similar stellar parameters, to try to emulate the effects of giant convection cells and pulsation, for a sample of models with eight different stellar parameter combinations. 
The resulting atmospheric dynamics are investigated, and compared to the standard DARWIN (Dynamic Atmosphere and
Radiation-driven Wind models based on Implicit Numerics) model results. 
Both approaches describe M-type AGB stars, with oxygen-dominated atmospheric chemistry.

\section{Modelling methods}

   \begin{figure}
   \centering
   \includegraphics[width=\hsize]{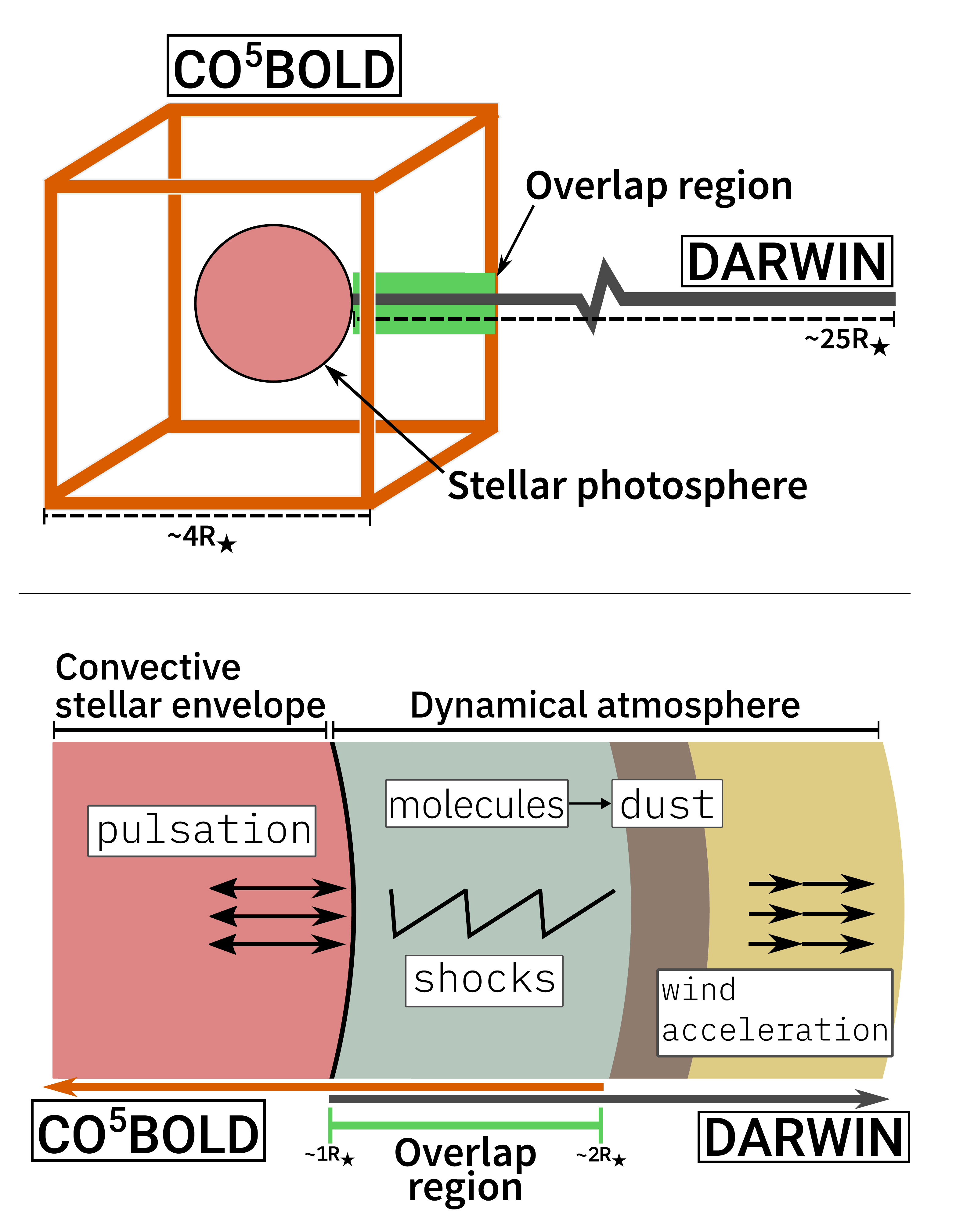}
      \caption{\textit{Top -} Schematic diagram showing an AGB star, the range of the DARWIN code (grey) and the outer boundary of the CO$^5$BOLD code (orange). As the CO$^5$BOLD models are star-in-a-box simulations, the computational box encases the full star. The region where the two modelling methods overlap is indicated in green.
      \textit{Bottom} - The physical processes taking place in different regions of the star. The pulsation of the stellar interior (red) triggers shockwaves in the lower atmosphere, where molecules form (green). The material is levitated to distances where dust can condense from the molecules (brown), which is radiatively accelerated outwards inducing a wind (yellow). The overlap region is again indicated in green.
      }
         \label{fig:overlap}
   \end{figure}

The two regions of great importance for the mass-loss mechanism are the convective stellar envelope, where pulsation originates, and the dynamical atmosphere, with dust formation and wind-driving. 
These two regions represent vastly different physical regimes, and modelling tends to focus on only one region (see Fig. \ref{fig:overlap}).

Here we use the CO$^5$BOLD radiation hydrodynamical code to simulate the interior of the star and the inner atmosphere. 
The CO$^5$BOLD models are 3D star-in-a-box models where the full star is described. 
The main purpose with the CO$^5$BOLD code is to describe the dynamics of the stellar envelope, which is a region dominated by convective flows. 
These models develop pulsations that have been shown to be realistic \citep{freytag_global_2017}.
However, due to computational constraints, the models only reach as far as $\sim 2R_\star$, making the approximate size of the
computational box $4 \times 4 \times 4$ R$_\star$. 
This outer boundary of a CO$^5$BOLD model is indicated by  orange in the upper panel of Fig. \ref{fig:overlap}.  
No dust formation is included in the code, so while the interior and the pulsations are well modelled, the wind-driving process is not described. 
The output of this code is time-series of 3D snap-shots of the structure of the star.

The dynamical atmosphere is described using the DARWIN code that includes frequency-dependent radiation hydrodynamics and time-dependent growth of dust.
The DARWIN code describes complex processes, such as shockwaves and dust formation that take place in the atmosphere, and predicts wind properties for different stellar parameters. 
The DARWIN models range from just below the photosphere at $\sim 0.9R_\star$ with the outer boundary typically located around $\sim 25R_\star$, where the wind has reached its final velocity.
The variability of the AGB star is typically an external input in DARWIN, simulated by variations of the luminosity and the gas velocity at the inner boundary.
The outputs of these models are dynamical properties, such as mass-loss rates and wind velocities, and radial snapshots of the atmospheric structures \citep[see][]{bladh_exploring_2015,hofner_dynamic_2016-1}.

\subsection{CO$^5$BOLD models}

The convective interior of the  AGB star is simulated using the CO$^5$BOLD  code,  a radiation-hydrodynamics (RHD) code that produces time-dependent 3D star-in-a-box models of the interior and lower atmosphere. 
This method of simulating AGB stars was first introduced in \cite{freytag_three-dimensional_2008}, where a few test models are explored. 
After improvements to the code \citep[outlined in][]{freytag_advances_2013}, a set of models with different stellar parameters was calculated and investigated in \cite{freytag_global_2017}. 
This set of models is also used in this paper. 

The start models for the first CO$^5$BOLD AGB star simulations were evolved from a hydrostatic 1D model \citep{freytag_three-dimensional_2008}. 
All the models presented in \cite{freytag_global_2017} are then produced from previous models by incrementally changing the luminosity at the core, envelope mass, and core mass. 

The CO$^5$BOLD code solves the coupled equations of compressible hydrodynamics and non-local grey radiation transport on a cartesian grid. 
Solar abundances are used, and assumed to be pertinent for M-type AGB stars, with tabulated grey opacities. 
The equation of state is similarly tabulated, and takes the ionisation of hydrogen and helium as well as H$_2$ formation into account. 

To describe gravity, a fixed gravitational potential with the form $1/r$ is applied.
This represents a central point mass, which is a good representation for the structure of an AGB star with a compact core surrounded by an extended stellar envelope. 
The central part of the gravitational potential is smoothened to avoid a singularity at the centre. 
For a description of the technical details, see  \cite{freytag_three-dimensional_2008,freytag_simulations_2012,freytag_global_2017}.

The grid cells and the computational box are all cubic, and all the outer boundaries of the computational box are open to both matter and to radiation \citep{freytag_boundary_2017}. 
The grid cells at the core are too coarse to resolve the zones where nuclear reactions take place. 
Instead an inner boundary is placed at $78R_\odot$, where constant radiation energy, corresponding to the luminosity of the star, is fed to the system. 
The velocity in the core region is damped to suppress dipolar flow through the pre-white dwarf centre.

As explored in \cite{freytag_global_2017}, these models develop radial and non-radial pulsations.
While the interaction of pulsation and convection certainly plays a major role, the exact mechanism behind this behaviour is not yet fully understood.
While the potential interaction between the pulsations and the damped core need more investigation, the pulsation properties extracted have been shown to be similar to observations \citep[see Fig. 9 in][]{freytag_global_2017}.

\subsection{DARWIN models}
The dynamical atmospheres of AGB stars, with pulsation-induced shocks and dust-driven winds, are simulated using the DARWIN  code.
Such models reach from just below the stellar photosphere ($\sim1R_\star$), out to beyond the wind acceleration region, where the wind has reached its terminal velocity $v_\infty$ ($\sim25R_\star$).

The DARWIN simulations start from hydrostatic dust-free atmosphere models, defined by the fundamental stellar parameters: effective temperature, luminosity, mass, and chemical composition.
A variation at the inner boundary is introduced in order to simulate the pulsation of the star, and is then gradually ramped up in amplitude. 
At every time-step the gas and dust dynamics are solved using frequency-dependent radiation hydrodynamics assuming spherical symmetry.
With the ramping of the variation at the inner boundary the models develop from hydrostatic into dynamical structures.
The modelling time is usually several hundred years, with the first 20-40 pulsation periods being the ramping period.

Solar abundances are assumed, and the wind-driving dust species consist of 0.1-1\micron sized silicate grains, here Mg$_2$SiO$_4$.
Silicate dust grains of these sizes have been observed around several M-type AGB stars \citep[see e.g.][]{norris_close_2012,ohnaka_clumpy_2016,ohnaka_clumpy_2017}.
The wind is driven by photon scattering on the dust particles, which accelerates the dust outwards. 
Momentum from the dust is transferred over to the gas through collisions, which induces a wind. 
In the present models we assume no drift between the dust and the gas. 

Dust growth and dust evaporation are time-dependent processes in the DARWIN models, and are simulated using the methods developed in \cite{gail_mineral_1999}.
Silicate dust (Mg$_2$SiO$_4$) condenses on the seed particles through reactions involving magnesium (Mg), silicon monoxide (SiO), and water (H$_2$O), with the growth rate limited by the available magnesium (Mg).
The dust nucleation process, that is, the formation of the first tiny solid, is poorly understood in oxygen-rich stellar environments \citep[see][]{gail_silicate_2016,gobrecht_dust_2016}, so dust seed particles, consisting of 1000 monomers, are assumed to exist.
Their growth to full-sized dust grains in regions with favourable conditions is described with time-dependent equations \citep{hofner_dynamic_2016-1}.
The dust condensation temperature is thus set by the microphysical properties of the dust. 
The grain growth rates on the other hand depend on densities, which in turn depend on the shockwave properties and the gas dynamics.
The seed particle abundance ($n_\mathrm{gr}/n_\mathrm{H}$) is a free parameter, defined as the ratio between number density of the seeds and the number density of the hydrogen atoms.  Based on the findings of \cite{bladh_exploring_2015}, $n_\mathrm{gr}/n_\mathrm{H}$ is set to $3\times10^{-15}$ for all the DARWIN models in this paper.

The system of partial differential equations describing the gas and dust dynamics, the dust growth and evaporation, and the radiative transfer is solved simultaneously for every time-step using a Newton-Raphson scheme \cite[for more details about the numerics see][]{hofner_dynamic_2016-1}. 
The spatial grid in the model is adaptive, taking the density and temperature gradients into account.

For a more detailed description of the DARWIN models see \cite{hofner_dynamic_2016-1}. 

\subsection{Inner boundary of DARWIN models}
\label{sect:ib}

The pulsation of the stellar surface layers is essential for the wind driving, as shockwaves triggered by the pulsations levitate material to distances with low enough temperatures for dust to be able to form. 
The DARWIN code does not model the interior of the star, where pulsations are excited. The effects of the variability (radial expansion and contraction of the surface layers and variation of the luminosity) are instead described in a parametric fashion. 
The standard approach to place a sinusoidal piston at the inner boundary of atmospheric AGB models was introduced by \cite{bowen_dynamical_1988}, and later adopted for the DARWIN models.
The form used for the velocity variation is
\begin{equation}
\label{eqn11}
u_\mathrm{in}(t) = \Delta u_p \cos\left ( \frac{2 \pi }{P} t \right)\enspace ,
\end{equation}
where $\Delta u_p$ is the velocity amplitude and $P$ is the pulsation period. 
From this we get a radial variation as
\begin{equation}
\label{eqn1}
R_\mathrm{in}(t) = R_0 + \frac{\Delta u_p P}{2 \pi} \sin{\left ( \frac{2 \pi }{P} t \right)} \enspace ,
\end{equation}
which describes the radial expansion and contraction of the impermeable inner boundary of the DARWIN models.

When assuming a fixed radiative flux at the inner boundary, the luminosity will vary in time proportionally to the square of the radius as $L_\mathrm{in} \propto R^2_\mathrm{in}$. 
This however was found to result in an overly small bolometric variation \citep[see][]{gautschy-loidl_dynamic_2004}.
To better match the flux variation, a free parameter $f_L$ was introduced to adjust the amplitude of the luminosity without changing the velocity amplitude. 
The luminosity variation can then be written as 
\begin{equation}
\label{eqn2}
\Delta L_\mathrm{in}(t) = L_\mathrm{in} - L_0= f_L \left (\frac{R^2_{in}(t) - R^2_0}{R^2_0} \right ) \times L_0
.\end{equation}

This approach has only two free parameters ($\Delta u_p$ and $f_L$).
However, it assumes sinusoidal shapes of the luminosity variation and the radial variation, constant amplitudes, and that luminosity and radial variations are locked in phase. 
In previous studies where the inner boundary was investigated, it was found that introducing a phase shift or changing the shapes or the amplitudes of the inner boundary have consequences for the resulting atmospheric structure and the dynamics \citep[see][]{liljegren_dust-driven_2016,liljegren_pulsation-induced_2017}. 

To be able to describe different dependencies on time of both the luminosity and the radial variation, the strictly sinusoidal description of the boundary previously used (Eq. \ref{eqn1} and \ref{eqn2}) can be replaced by a Fourier description as

\begin{equation}
\label{eqn3}
R_\mathrm{in}(t) = R_0 \sum_i  A_i \sin((2\pi/t_\mathrm{tot})  i  t - \delta_i)
,\end{equation}

\begin{equation}
\label{eqn4}
L_\mathrm{in}(t) = L_0 \sum_i  A_i \sin((2\pi/(t_\mathrm{tot}))  i  t - \delta_i)
.\end{equation}

Results from pulsation models can then be used to describe the inner boundary instead of the previous, parameterised approach.
The corresponding Fourier components $A_i$ and $\delta_i$ used in this paper are derived by fast Fourier transform based on luminosity and radial variations extracted from the CO$^5$BOLD models.

\section{Comparing the 1D and 3D atmospheres}
\label{sect:comp}

\label{morph}
    \begin{figure}
   \centering
   \includegraphics[width=\hsize]{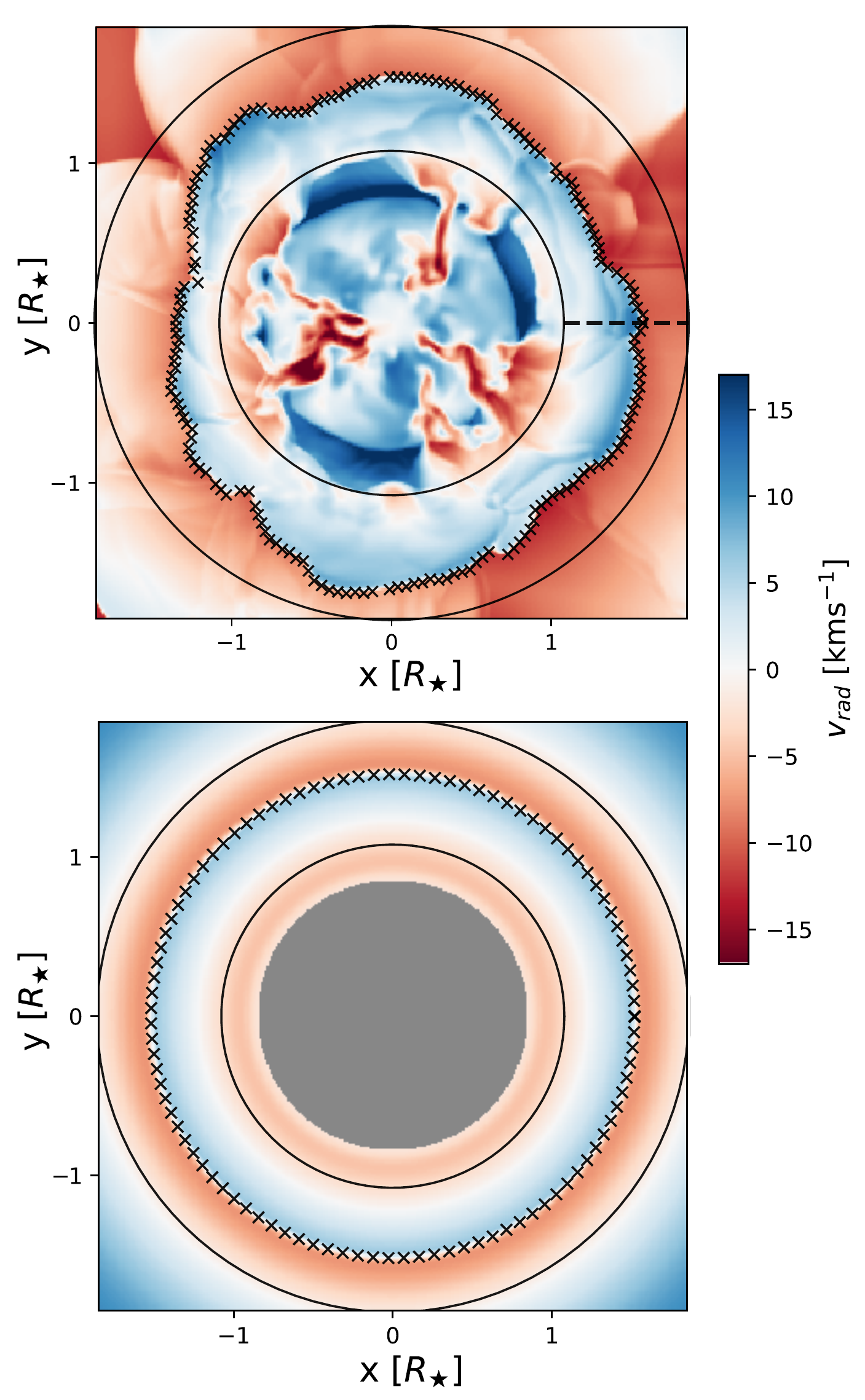}
      \caption{The cross-section of a snapshot of the two models, showing the radial velocity. 
      The black circles represent $R=1.1R_\star$ and $R=1.9R_\star$, the region discussed in Sect.~\ref{sect:amorph}, and the crosses trace the shockwaves of the models at $\sim1.5R_\star$. 
      \textit{Top} - CO$^5$BOLD model \textit{st28gm06n26}, with stellar parameters $1M_\odot$, $T_\textrm{eff} = 2737$K and $L_\star = 6955L_\odot$. The dashed line indicates the x-direction of the model. Gas movements in this direction are later used to illustrate the local dynamics in the 3D models. 
      \textit{Bottom} - The corresponding snap-shot for the DARWIN model \textit{l70t27}, with stellar parameters $1M_\odot$, $T_\textrm{eff} = 2700$K and $L_\star = 7000L_\odot$. 
      Beyond $\sim 1.9R_\star$ the DARWIN models show the onset of wind  (positive velocities), in contrast to the windless CO$^5$BOLD model.}
         \label{fig:morph}
   \end{figure}

\begin{table*}[]
\begin{center}
\caption{Models used for comparison.}
\label{table:comp}
\begin{tabular}{l|llll|l}

Name                      & Code       & Base model           & $L_\star$ $[L_\odot]$ & $T_{\textrm{eff}}$ [K]  & Notes                                                                                                                                                                                                      \\ \hline
CO$^5$BOLD mean  & CO$^5$BOLD & \textit{st28gm06n26} & 6955                  & \multicolumn{1}{l|}{2737}                      & Mean structure of a 3D model.\\
CO$^5$BOLD x-dir & CO$^5$BOLD & \textit{st28gm06n26} & 6955                  & \multicolumn{1}{l|}{2737}                      & Local structure of a 3D model.                          \\
DARWIN u2        & DARWIN     & \textit{l70t27}      & 7000                  & \multicolumn{1}{l|}{2700}                      & $\Delta u_p=2$ (from Eq. \ref{eqn11} and \ref{eqn1})                                                                                                                 \\
DARWIN u4        & DARWIN     & \textit{l70t27}      & 7000                  & \multicolumn{1}{l|}{2700}                      & $\Delta u_p=4$ (from Eq. \ref{eqn11} and \ref{eqn1})                                                                                                                          
\end{tabular}
\end{center}
{Two ways are used to extract radial velocities and luminosity from a 3D model, either by considering the mean behaviour of the structure or the local behaviour (see Sect. \ref{sect:comp} for more detailed explanation). This is compared to two DARWIN models, with the same stellar parameters but different inner boundary conditions. }
\end{table*}

As mentioned, the CO$^5$BOLD models reach only into the lower atmosphere, out to around $\sim2R_\star$, while the spatial range of the DARWIN models is from $\sim1R_\star$ out to $\sim25R_\star$. 
The two modelling methods therefore both describe the region in the range $\sim1-2R_\star$, which is where the shocks created by the pulsation develop but is before any significant amount of dust is formed (see Fig. \ref{fig:overlap}). 
The gas layers here will follow more or less ballistic trajectories, and the amplitude and occurrence of the shocks are highly dependent on the pulsation properties of the surface layers of the star.
The density, timing, and velocity of the shockwaves will also affect how much matter is levitated to distances where dust can condense, and therefore how much dust forms and how effective the wind acceleration is. 

In this lower atmosphere region, where the models overlap, we compare the gas dynamics in the DARWIN models, which are a consequence of the inner boundary condition, to the gas dynamics of the CO$^5$BOLD models, where the shockwaves triggered by pulsations emerge in the simulations.

Throughout this section the CO$^5$BOLD model \textit{st28gm06n26}, with the stellar parameters $1M_\odot$, $T_\textrm{eff} = 2737$K and $L_\star = 6955L_\odot$, is used as an example for the 3D models and compared to two corresponding DARWIN models \textit{l70t27u2/u4}, with stellar parameters $1M_\odot$, $T_\textrm{eff} = 2700$K and $L_\star = 7000L_\odot$, and with either $\Delta u_p=2$ or $4$ (from Eq. \ref{eqn11} and \ref{eqn1}). 
The CO$^5$BOLD model \textit{st28gm06n26} has previously been used as a standard model \citep[see][]{freytag_global_2017} to showcase the behaviour of the 3D models.
The stellar parameters ($1M_\odot$, $T_\textrm{eff} \sim 2700$K, $L_\star \sim 7000L_\odot$) are typical for an AGB star, and the  \textit{st28gm06n26} model reproduces observations of pulsation period against luminosity very well \citep[left panel of Fig. 9 in][]{freytag_global_2017}. 
\textit{l70t27u2/u4} are the corresponding DARWIN models, with similar stellar parameters.

For the CO$^5$BOLD model, the luminosity and the radial velocities are derived in two ways: either by considering the mean structure, indicated in orange in all following figures, or only the $+x$-direction, shown in green. 

The output of the CO$^5$BOLD models are 3D snapshots of the structure, at each time-step. 
For the mean structure radial velocity, the radial velocity is averaged over spherical shells (same radial distance) for each of these snapshots. 
This quantity then represents the net gas movement of the models at a specific radial distance and time-step. 
If the mean radial velocity is positive, then the gas is on average moving outwards at this distance, and vice versa for a negative velocity.
The luminosity for the mean structure is the average luminosity flowing out from all sides of the computational box. 

For the one-direction scheme, the radial velocity and luminosity in the $+x$-direction is used. 
The one-direction radial velocity represents the local gas movement, in contrast to the mean structure. 
Throughout this section the mean structure properties are referred to as the CO$^5$BOLD mean (represented by orange), and the $+x$-direction properties as  the CO$^5$BOLD x-dir (represented by green). 
This allows us to compare both the mean behaviour and the behaviour in only one direction for the 3D model, with the DARWIN model with either $\Delta u_p=2$ or $4$, referred to as DARWIN u2 or DARWIN u4, shown in grey and black, respectively. Table \ref{table:comp} shows an overview of the models used in this section.

The comparison is divided into two parts: in Sect. \ref{sect:amorph} we look at snapshots in time and compare the spatial differences between the models, while in Sect. \ref{cc} the temporal variations are investigated.

\subsection{Results - Atmosphere morphology}
\label{sect:amorph}

\begin{figure*}
\centering
\includegraphics[width=\hsize]{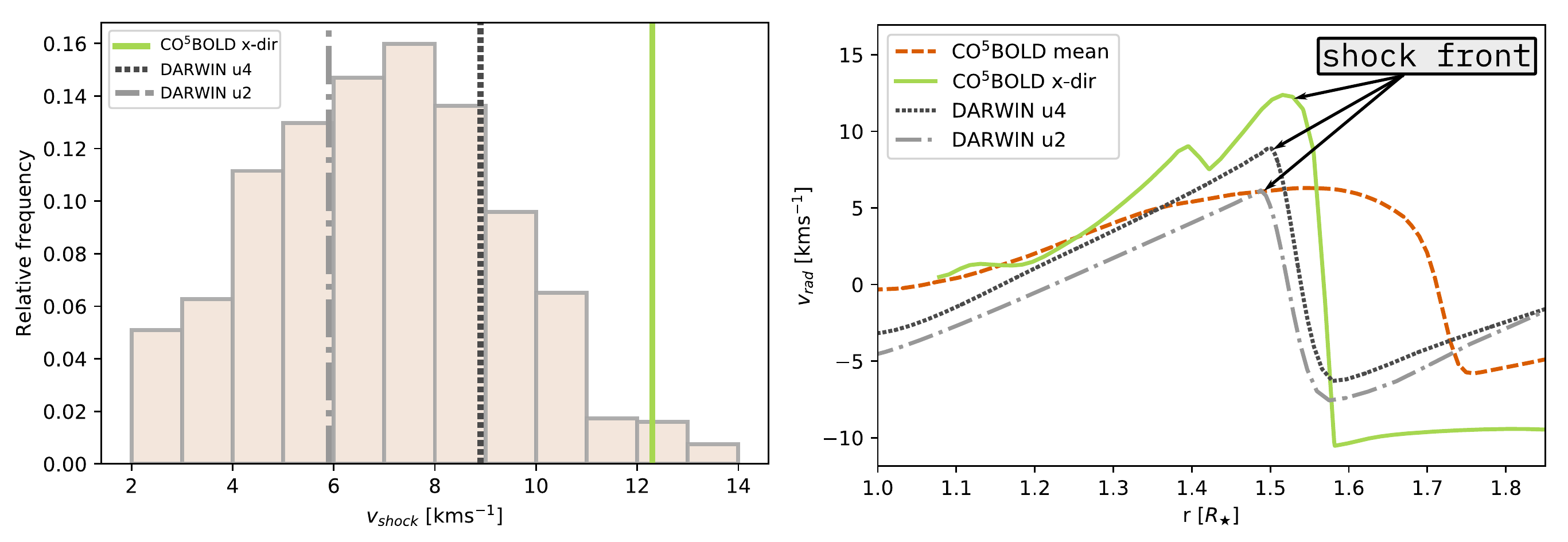}
\caption{\textit{Left - } The distribution of maximum radial gas velocities in the full 3D shockwave of the CO$^5$BOLD model \textit{st28gm06n26}, the same snapshot as shown in the top panel of Fig. \ref{fig:morph}. The lines indicate the corresponding radial velocities for the DARWIN model \textit{l70t27} for $u_p=2$ and $4$ (grey and black), and the maximum velocity in one direction in the CO$^5$BOLD model \textit{st28gm06n26} (green here, the direction is indicated by black dashed line in the top panel of Fig. \ref{fig:morph}). \textit{Right - } Radial velocity against the radial distance, for the different models. The shock front of the CO$^5$BOLD \textit{x}-dir, DARWIN u2, and DARWIN u4 are indicated. There is no clear shock front present for the CO$^5$BOLD mean.}
\label{fig:veld}
\end{figure*}

\subsubsection{Snapshots of the radial velocity}
Figure \ref{fig:morph} shows the radial-velocity cross-section of one snapshot for a CO$^5$BOLD model in the top panel and a DARWIN model in the bottom panel. 
The black circles represent $R=1.1R_\star$ and $R=1.9R_\star$, which encase the atmospheric region discussed in detail in this section. 
The shock fronts in both the CO$^5$BOLD model and the DARWIN model are at $R\sim1.5R_\star$ for the picked snapshots, and traced by determining where the gas has a sharp velocity inversion in this interval (indicated by black crosses).

There are certain similarities between the two models. 
The amplitudes of the velocities are comparable in both the shockwave and the in-falling material for the models. 
However with 1D models comes the inherent assumption of spherical symmetry. 
As seen for the CO$^5$BOLD model, this is not necessarily true. 
While the shock in the 3D models is global in scale, as it covers most of the surface of the star, the maximum velocity reached by the gas in the shock front is not uniform. 

This can be seen in the left panel of Fig. \ref{fig:veld}, which shows the distribution of the maximum gas velocities in the radial direction of the shockwave for the CO$^5$BOLD  snapshot shown in Fig.~\ref{fig:morph}. 
The distribution is derived from a snapshot of the 3D structure (the cross-section seen in Fig. \ref{fig:morph} is from the same snapshot). 
While the distribution has a peak at $v=7-8$km s$^{-1}$, there is a broad spread in velocities, with the maximum one being $>13$km s$^{-1}$. 
The lines in the plot indicate the maximum gas velocity for the $+x$-direction of the 3D snapshot (green), and the two DARWIN models (grey and black). 

This means that in some directions, where the radial velocities are low, not enough material is levitated to distances where dust can form.
Dust should therefore not form in a uniform spherical layer around the star, but rather in a clumpy formation with dust in some directions and no dust in other directions \citep[see][]{freytag_three-dimensional_2008}. 
Further, in directions with little or no dust the radiation pressure cannot overcome the gravitational well.

The right panel of Fig. \ref{fig:veld} shows the radial velocity with distance from the centre of the star, for both models at the same time-step as shown in Fig. \ref{fig:morph}. 
The radial velocity for the mean structure is shown in orange, and the radial velocity in the $+x$-direction is indicated by green, with the two DARWIN models in grey and black. 
The DARWIN models and the CO$^5$BOLD model in $+x$-direction have a clear shock structure present, with sharp shock fronts. 
In the CO$^5$BOLD mean model the radial-velocity curve is smeared with no clear shock front, as the shock front has different velocities in different directions, as seen in the upper panel of Fig. \ref{fig:morph} and in the left panel of Fig. \ref{fig:veld}. 

The DARWIN models should therefore not be compared to mean velocity dynamics of the 3D models for the lower atmosphere region, but rather to the dynamics in only direction.

\subsubsection{Surface area covered by shockwaves}
\label{sect:shock}

    \begin{figure}
   \centering
   \includegraphics[width=\hsize]{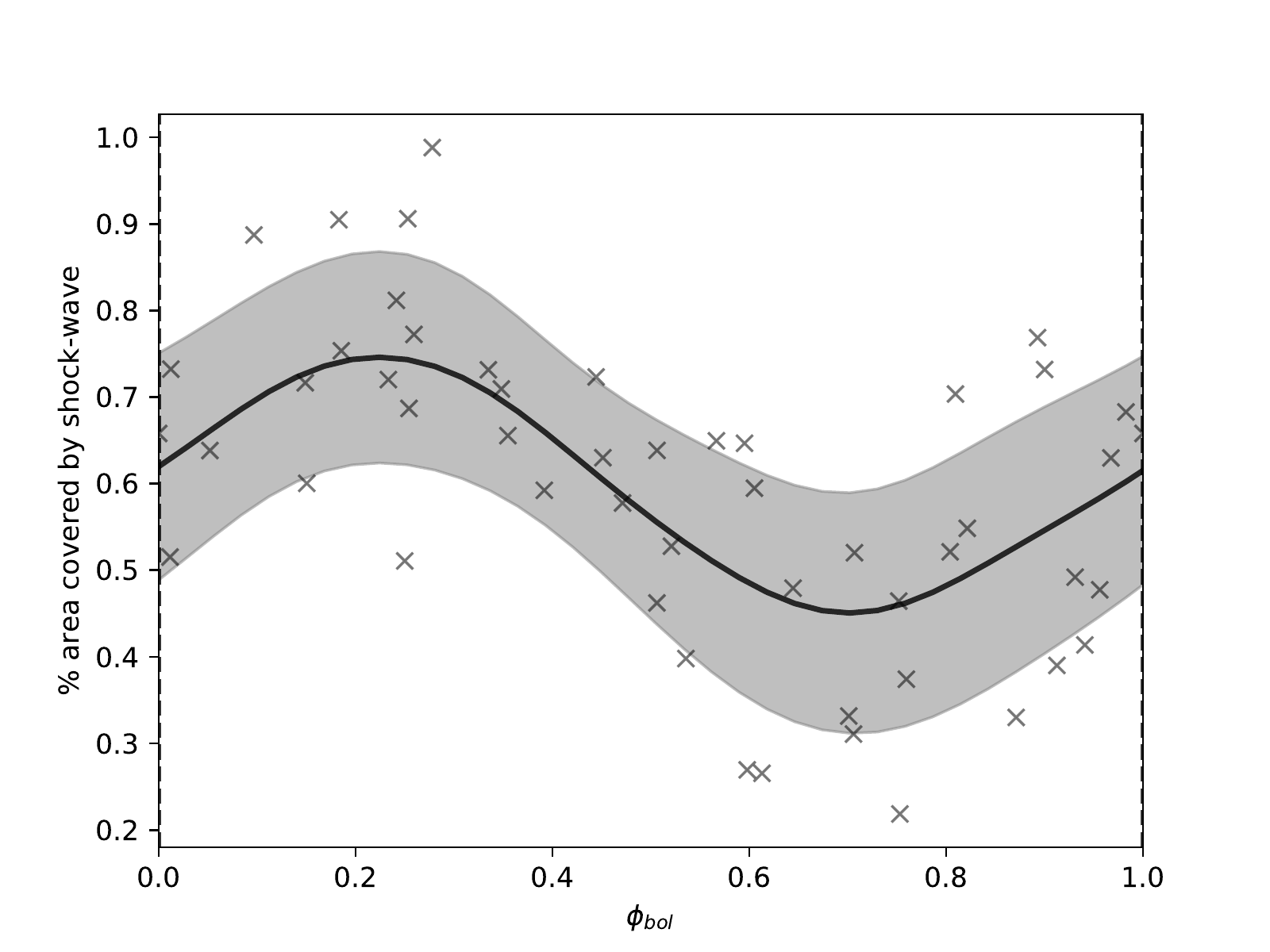}
      \caption{The grey crosses are the area of the surface covered by a shockwave, for $\sim50$ snapshots of the CO$^5$BOLD model \textit{st28gm06n26} at different bolometric phases ($\phi_{bol}$). The black solid line is a running mean, with an interval of 0.1 in $\phi_{bol}$, and the grey area is the corresponding standard deviation. }
         \label{fig:area}
   \end{figure}

In the spherically symmetric DARWIN models, shockwaves always cover $100\%$ of the surface. 
This is not the case for the 3D models, and can be seen in Fig. \ref{fig:area}, which shows the percentage of the surface of the 3D model that is covered by shockwaves with the bolometric phase $\phi_{bol}$, for around 50 3D structure snap-shots at different phases.
The shockwave was traced similarly to Fig. \ref{fig:morph}, by looking for sharp velocity inversions in the radial-velocity field in the inner atmosphere region between $R=1.1R_\star$ and $R=1.9R_\star$.
The black line shows the running mean with a window of 0.1 in  $\phi_{bol}$, and the grey area is the corresponding standard deviation.  
The shockwave coverage seems to peak on average at around $70\%$ during a cycle for this model. 

In the direction with no shockwave, no material will be levitated and no dust formed during that pulsation cycle. 
When assuming total shockwave coverage, the DARWIN models might therefore overestimate the mass-loss rates.

\subsection{Results:  cycle-to-cycle variation}
\label{cc}

In addition to the inherent non-spherical morphology demonstrated by the 3D models for the lower atmosphere, there is cycle-to-cycle variation present in the CO$^5$BOLD models.

\subsubsection{Luminosity and radius}

   \begin{figure}
   \centering
   \includegraphics[width=\hsize]{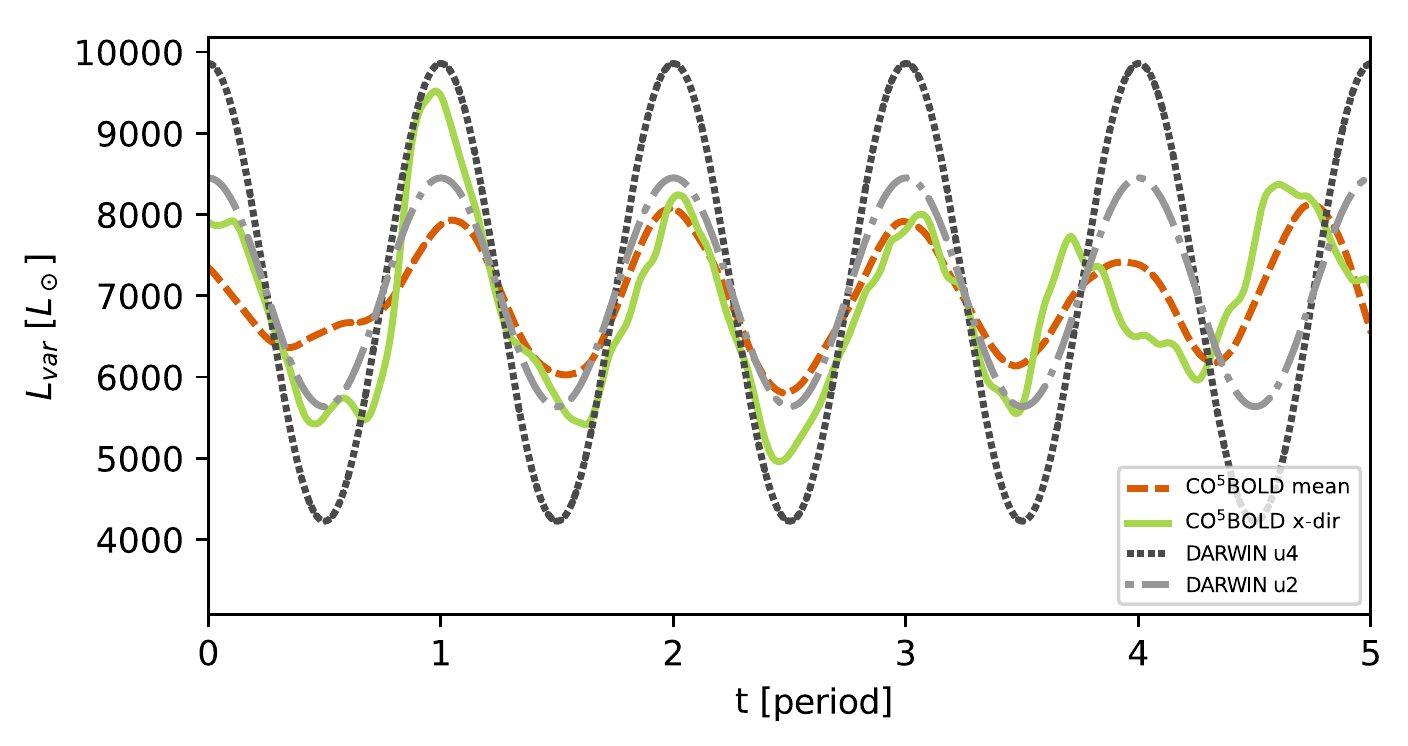}
   \includegraphics[width=\hsize]{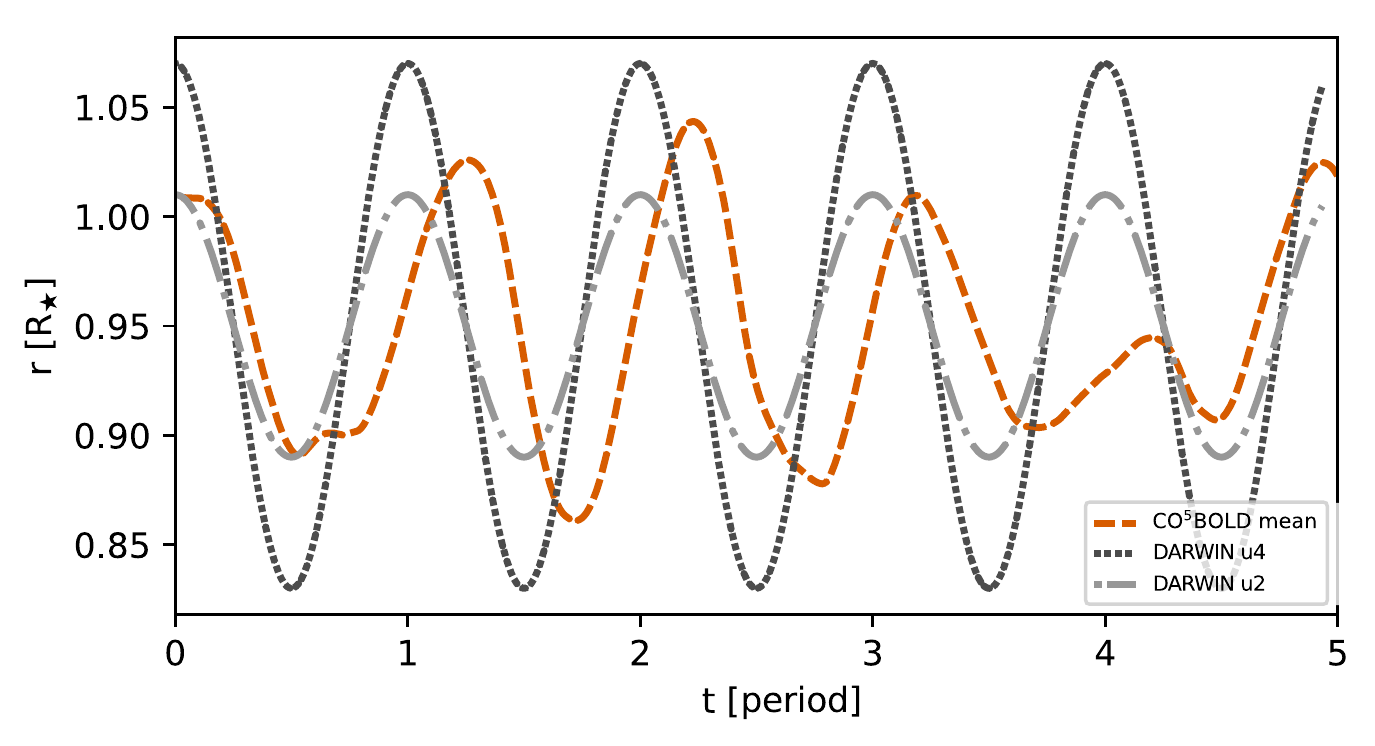}
      \caption{\textit{Top - }Comparison of the luminosity variation of the CO$^5$BOLD model \textit{st28gm06n26}, both averaged over all directions (orange) and only in the +x direction (green), as well as the luminosity variation of the DARWIN \textit{l70t27} $u_p=2$km s$^{-1}$ and 4km s$^{-1}$ models (grey and black).
      \textit{Bottom -} Comparison of the inner-most mass shells, representing the inner radial boundary,  of the DARWIN \textit{l70t27} $u_p=2$km s$^{-1}$ and 4km s$^{-1}$ models (grey and black) and the mass shell at a comparable distance from the CO$^5$BOLD model \textit{st28gm06n26}, averaged over all directions (orange). This is the same time-interval as the top panel of Fig.~\ref{fig:rvcomp}.}
         \label{fig:lvar}
   \end{figure}

\begin{figure*}
\centering
\includegraphics[width=\hsize]{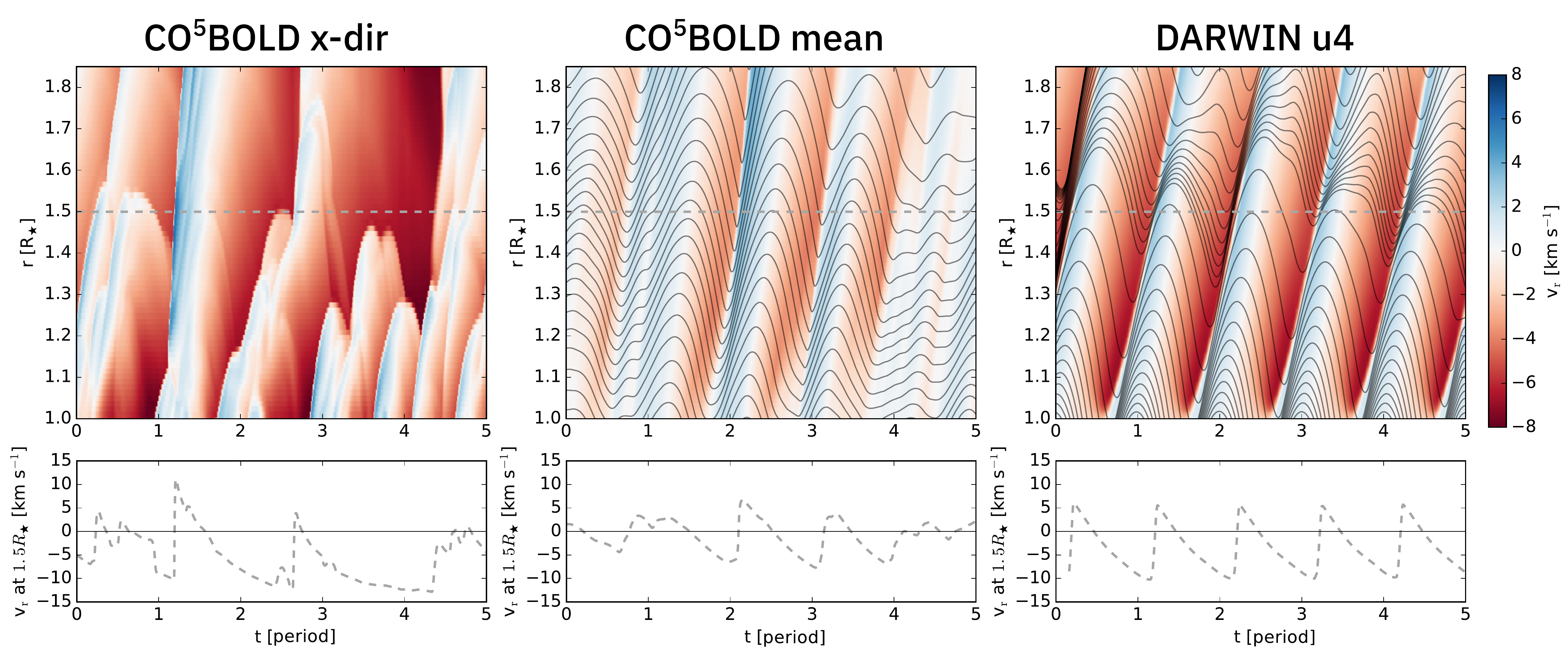}
\caption{The upper panels show the radial velocities, and mass shells where applicable (grey lines in middle and right panel). The lower panels show the radial velocity $v_r$ at $R=1.5R_\star$, which is drawn in dashed grey lines in the upper panel. This is the same time interval as shown for the luminosity in Fig.~\ref{fig:lvar}. \textit{Left - }CO$^5$BOLD model \textit{st28gm06n26}, considering only the +x direction. \textit{Middle -} CO$^5$BOLD model \textit{st28gm06n26} spherical averages. \textit{Right -} DARWIN \textit{l70t27} u4 model.  }
\label{fig:rvcomp}
\end{figure*}

The top panel of Fig. \ref{fig:lvar} shows the luminosity variation for the DARWIN model \textit{l70t27}, with $\Delta u_p=2$ km s$^{-1}$ and $\Delta u_p=4$ km s$^{-1}$, and the luminosity variation for CO$^5$BOLD model \textit{st28gm06n26}, both spatially averaged over all sides of the computational box and only in the $+x$-direction, against the period. 
The luminosity variation of the DARWIN models have, by design, amplitude and period with no differences between the cycles. 
For the CO$^5$BOLD model both the mean luminosity and luminosity in the $+x$-direction also show clear periodicity, with amplitudes similar to the DARWIN u2 model.
The CO$^5$BOLD x-dir model also reaches higher luminosity amplitudes during cycle 1, up to amplitudes similar to that of the DARWIN u4 model.

There is however a cycle-to-cycle variation for both amplitudes and cycle periods, for both the mean and the x-dir CO$^5$BOLD light curves. 
For the CO$^5$BOLD mean light curve, this variation is relatively small, and mostly visible in cycles four and five. 
For the fourth cycle, the amplitude of the luminosity drops quite significantly, by almost $50\%$. 
The subsequent luminosity peak then occurs earlier. 

For the CO$^5$BOLD x-dir light curve, this irregular behaviour is more prominent than for the mean luminosity. 
As seen in the top panel of Fig. \ref{fig:lvar} the amplitude almost doubles during the first cycle, compared to cycles two and three.
During cycle four there is most likely a convective event interrupting the regular pulsation in this direction, and again the subsequent luminosity peak occurring at earlier time.

The behaviour of the radius seems to be correlated to the behaviour of the luminosity in the 3D models. 
The bottom panel of Fig.~\ref{fig:lvar} shows the inner-most mass shell of the DARWIN \textit{l70t27} u2 and u4 models (grey and black), as well as the comparable mass shell from the spherically averaged CO$^5$BOLD model \textit{st28gm06n26} (orange).
The trend is similar to the luminosity variation; when the luminosity amplitude is larger the relative radial expansion of the model is also larger, and vice versa. 
Therefore, even though the CO$^5$BOLD model clearly has a periodic expansion and contraction of the radius, there are cycle-to-cycle variations. 
Both the amplitude of the radial variation and the timing of the maximum variation varies.

Just as the non-spherical morphology of the 3D model should affect the dust forming, and thus the outflow, in different directions, these temporal variations should affect conditions for the wind-driving, from cycle-to-cycle.
Changes in the luminosity amplitude influence both the temperature structure, which is important for the condensation distances of the dust, and the radiation pressure being exerted on the dust. 
Similarly, the differences in radial amplitude should affect the radial velocities in the stellar atmosphere and the amount of dust formed, and thus the wind driving.

\subsubsection{Radial velocity}
\label{sect:radvel}

There is a cycle-to-cycle variation in the radial velocity in the inner part of the atmosphere in the 3D models.
This can be seen in Fig. \ref{fig:rvcomp}, where the radial velocities for the three cases (3D models looking in the $+x$-direction, 3D model spherical mean, and a 1D model) are plotted. 
For the 3D mean case and the 1D model, the corresponding mass shells\footnote{Lagrangian mass shells, where each line is a spherical surface that contains a constant mass. This is used to show the mean gas movements with time at different depths in the atmosphere. } are over plotted.  
The lower panels show the radial velocity at $R=1.5R_\star$. 

The DARWIN model to the right shows very regular behaviour, a consequence of the periodic inner boundary condition. 
The shockwaves always occur with the same periodic interval, and the shock velocities at $R=1.5R_\star$ are the same from cycle-to-cycle.
About the same amount of material is levitated into the dust-forming region in each period and a similar amount of dust will be formed. 
The resulting outflow is therefore often very stable.

The general behaviour of the spherical means of the 3D model, shown in the middle of Fig. \ref{fig:rvcomp}, is similar to the 1D model. 
Shockwaves develop periodically, at mostly the same interval.
The radial-velocity amplitude at $R=1.5R_\star$ is also similar to the 1D case, around 5 km s$^{-1}$.
The shock fronts of the 3D mean model are not as sharp as in the case of the 1D model, which is a consequence of the non-spherical morphology of the shocks (discussed in Sect. \ref{morph}). 

The radial velocity in the $+x$-direction of the 3D models is shown to the left in Fig. \ref{fig:rvcomp}. 
The shapes of the shocks passing through $R=1.5R_\star$ are similar to the case of the 1D models, however the velocity amplitudes are vastly varying from 3 km s$^{-1}$ up to 10 km s$^{-1}$ in the interval shown. 
This again highlights the non-spherical character of the shockwaves that develop in the 3D models. 
There seems to be some periodicity, as shocks develop in the three first cycles.
However there is no shock that reaches $1.5R_\star$ in cycle 3.
 As mentioned in Sect. \ref{sect:shock} and seen in Fig. \ref{fig:area}, only about 70$\%$ of the surface of the star will be covered by the global shockwaves that develop in the inner region of the atmosphere ($R\sim1.1-2R_\star$).
 This is probably the reason why no shockwave propagates in the $+x$-direction during cycle 4.

\subsubsection{Phase shift}

Another interesting quantity shown to be important for modelling wind properties is the phase shift between the luminosity variation and the radial variation  \citep[see][]{liljegren_dust-driven_2016,liljegren_pulsation-induced_2017}.
This phase shift essentially measures the time between luminosity maximum and maximum extension of the radius. 
\cite{liljegren_pulsation-induced_2017} found that phase information could be deduced from observations of radial-velocity curves derived from second overtone CO-lines. 
The observed radial-velocity curves, which have very distinct S-shapes, amplitudes and line-doubling features, could successfully be reproduced by DARWIN models if a phase shift of $\sim 0.2$ periods was added. 

In the CO$^5$BOLD models investigated here we find that the maximum expansion of the radius occurs around $0.1-0.3$ periods  after the luminosity maximum, consistent with previous findings. 
An example of this can be seen by comparing the top  and bottom panels of Fig.~\ref{fig:lvar}, which show luminosity variation and radial variation for the same time interval.
The DARWIN models (grey and black) have maximum luminosity and maximum expansion occurring at the same time, as a consequence of the chosen boundary condition.
The maximum mean expansion of the CO$^5$BOLD model (orange) however consistently occurs after maximum mean luminosity.

\section{Estimating the effects on the wind properties}
\label{sect:darco}

\subsection{Different boundary condition schemes for DARWIN models}
\label{darco}
   \begin{figure}
   \centering
   \includegraphics[width=\hsize]{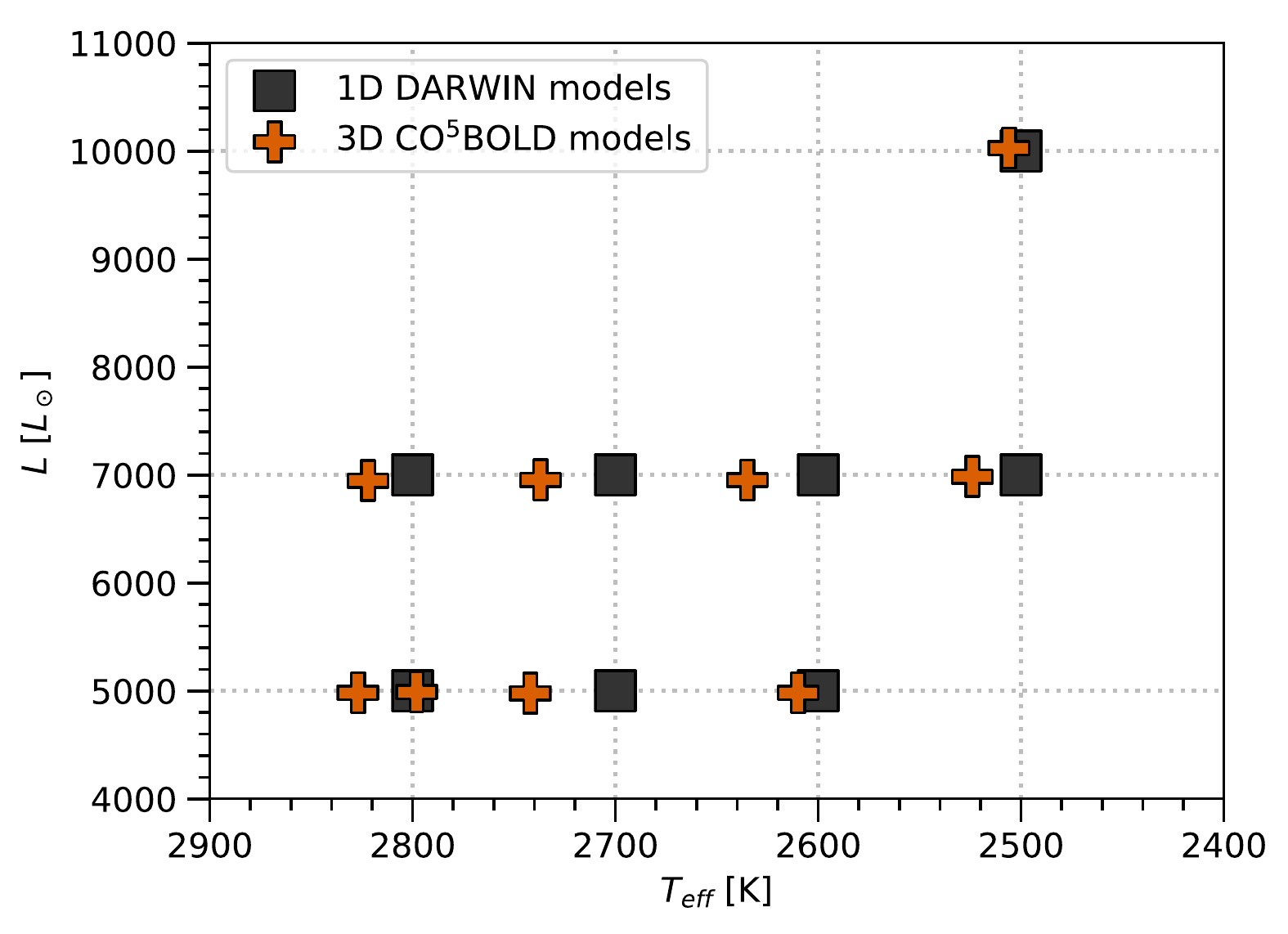}
      \caption{HR diagram showing the stellar parameters of the CO$^5$BOLD models \citep[from][orange crosses]{freytag_global_2017} and the corresponding DARWIN models \citep[based on models from][grey boxes]{bladh_exploring_2015}.}
         \label{fig:modov}
   \end{figure}

We try to imitate the complex physical conditions in the atmosphere inflicted by the pulsation and the giant convection cells, using the CO$^5$BOLD models to derive lower boundary conditions (shortened to BC in the following text) for DARWIN models of similar stellar parameters.
A first attempt to do this was made by \cite{freytag_three-dimensional_2008}, however only the radial movements $R_\textrm{in}(t)$ for two models were extracted, and $L_\textrm{in}(t) \propto R^2_\textrm{in}(t)$ was used to describe the variation of luminosity. 
The time spans covered by these 3D models were relatively short (8 and 14 years).
Furthermore the derived boundary conditions were only applied to wind models for C-rich AGB stars.

In the present work we try to improve the previous approach, with several advantages: firstly, we explore the systematic effects, using the grid of CO$^5$BOLD AGB models first introduced in \cite{freytag_global_2017} as a starting point. 
Secondly, the 3D models used here have longer simulation times, of around  25 years or more, which correspond to around 25 periods.
Thirdly, both the luminosity $L_\textrm{in}(t)$ and radial $R_\textrm{in}(t)$ variations are extracted independently from the 3D models, and used as inner boundary conditions. 
Boundary conditions are derived for each of the 3D models and applied to the matching DARWIN models for M-type AGB stars \citep[based on models in][]{bladh_exploring_2015}, with similar luminosity and effective temperature, shown in Fig. \ref{fig:modov}.

We used three different modelling approaches to investigate the effects on the wind properties: 
\begin{itemize}
\item \textbf{DARWIN u2 and u4} - Standard DARWIN models, using the boundary condition from Eqs. \ref{eqn1} and \ref{eqn2}, with $f_l = 2$, $\Delta u_p = 2$ km s$^{-1}$ for the u2 model, and $\Delta u_p = 4$ km s$^{-1}$ for the u4 model. Both the u2 and the u4 inner boundary conditions are combinations of parameters typically used \citep[see e.g.][]{eriksson_synthetic_2014, bladh_exploring_2015, hofner_dynamic_2016-1}. 
\item \textbf{DARWIN with CO$^5$BOLD mean BC} - Using the mean luminosity and the mean radius variation of the CO$^5$BOLD models as the DARWIN inner boundary condition (BC) $L_\mathrm{in}$ and $R_\mathrm{in}$ respectively. 
\item \textbf{DARWIN with CO$^5$BOLD x-dir BC} - Inner boundary condition derived from the CO$^5$BOLD models by considering the gas dynamics and luminosity in only one direction, which should emulate the local behaviour of the 3D models and sample the full distribution of velocities seen in Fig. \ref{fig:veld}.
\end{itemize}

Each DARWIN model, with the stellar parameters shown in Table \ref{tab1}, is simulated with the three different boundary condition approaches. 
The results from the two CO$^5$BOLD BC are compared to the standard DARWIN u2 and u4 results, to evaluate if there are any systematic effects of using boundary conditions derived from the 3D models. 

\begin{table*}[hbt]
\begin{center}
\caption{Model parameters.}
\label{tab1}
\begin{tabular}{|r|cccc|r|cccc}
CO$^5$BOLD   &                           &                           &                          &                               & DARWIN &                           &                           &                          &                               \\ \hline
Name         & $L_\star$ $[L_\odot]$ & $R_\star$ $[R_\odot]$ & $T_{\textrm{eff}}$ [K] & $P_{\textrm{puls}}$ [days] & Name   & $L_\star$ $[L_\odot]$ & $R_\star$ $[R_\odot]$ & $T_\textrm{eff}$ {[}K{]} & $P_\textrm{puls}$ {[}days{]} \\ \hline
\textit{st28gm07n001} & 10 028                     & 531                       & 2506                     & 820                           & \textit{l10t25} & 9979                      & 447                       & 2500                     & 629                           \\
\textit{st26gm07n002} & 6986                      & 437                       & 2524                     & 593                           & \textit{l70t25} & 7077                      & 388                       & 2500                     & 484                           \\
\textit{st26gm07n001} & 6953                      & 400                       & 2635                     & 516                           & \textit{l70t26} & 7077                      & 367                       & 2600                     & 484                           \\
\textbf{\textit{st28gm06n26}}  & \textbf{6955}                       & \textbf{371}                       & \textbf{2737}                     & \textbf{471  }                         & \textbf{\textit{l70t27}} & \textbf{7077}                      & \textbf{348}                       & \textbf{2700}                     & \textbf{484}  \\
\textit{st29gm06n001} & 6948                      & 348                       & 2822                     & 419                           & \textit{l70t28} & 7077                      & 328                       & 2800                     & 484                           \\
\textit{st27gm06n001} & 4982                      & 345                       & 2610                     & 448                           & \textit{l50t26} & 5010                      & 315                       & 2600                     & 373                           \\
\textit{st28gm05n002} & 4978                      & 313                       & 2742                     & 393                           & \textit{l50t27} & 5010                      & 297                       & 2700                     & 373                           \\
\textit{st28gm05n001} & 4990                      & 300                       & 2798                     & 374                           & \textit{l50t28} & 5010                      & 281                       & 2800                     & 373                           \\
\textit{st29gm04n001} & 4982                      & 294                       & 2827                     & 338                           & \textit{l50t28} & 5010                      & 281                       & 2800                     & 373                          
\end{tabular}
\end{center}
{Overview of the stellar properties of the grid of CO$^5$BOLD models \citep[from][]{freytag_global_2017}, from which boundary conditions are extracted, and the corresponding DARWIN models \citep[from][]{bladh_exploring_2015}. The model in bold font is the one used in the previous section.}
\end{table*}

   \begin{figure}
   \centering
   \includegraphics[width=\hsize]{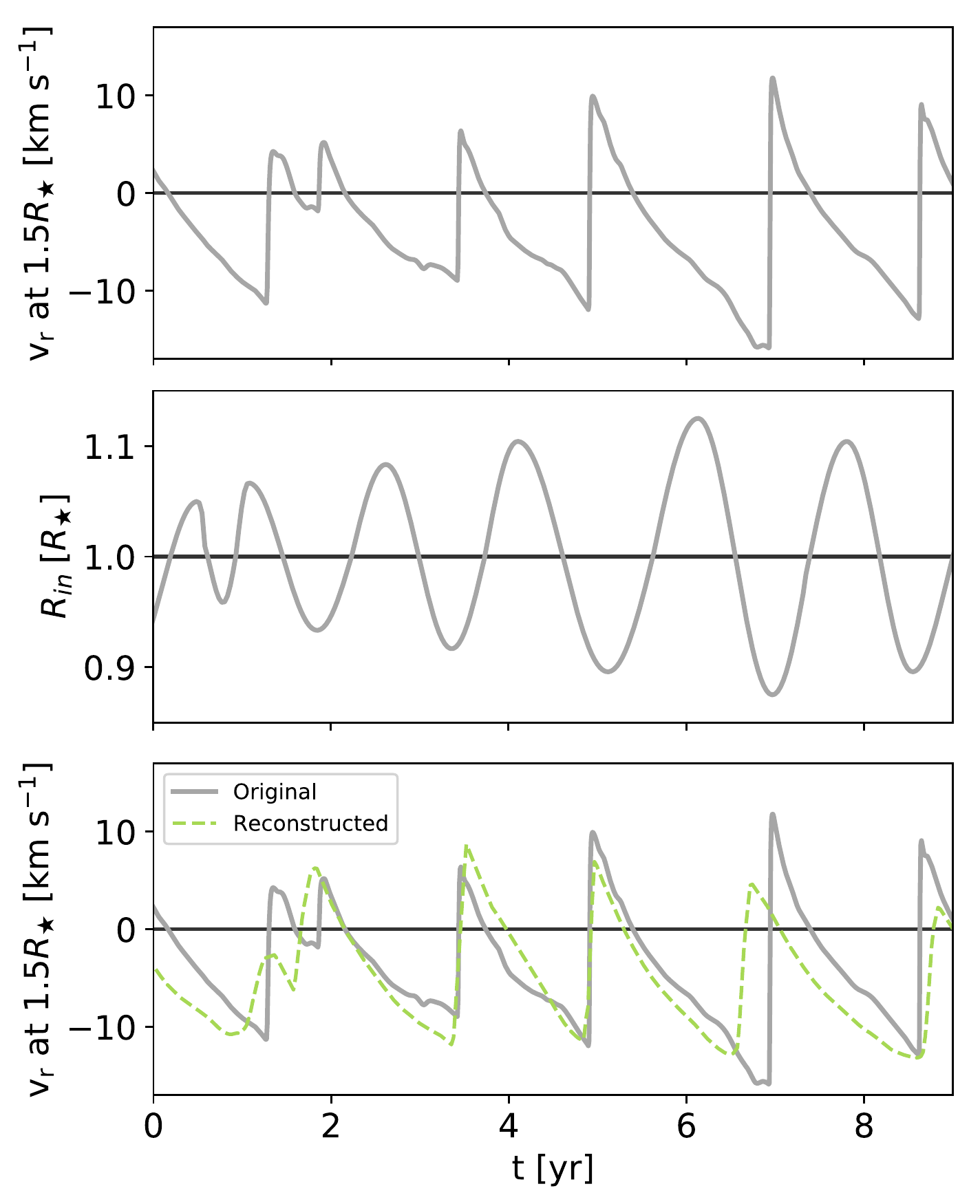}
      \caption{\textit{Top -} The velocity at $R = 1.5R_\star$, in the CO$^5$BOLD model \textit{st28gm06n26}, considering only the +x direction, the same quantity as in the lower right panel in Fig. \ref{fig:rvcomp}. 
      \textit{Middle -} The reconstructed boundary condition, used in the DARWIN model \textit{l70t27}. 
      \textit{Bottom -} The velocity at $R = 1.5R_\star$ for the DARWIN model \textit{l70t27} using the reconstructed boundary condition from the middle panel (green), compared to the velocity of the original CO$^5$BOLD model (grey). }
         \label{fig:recon}
   \end{figure}

For the CO$^5$BOLD mean BC, we use the luminosity averaged in all directions, with an example shown in orange in Fig. \ref{fig:lvar}.  
For the CO$^5$BOLD x-dir BC, the luminosity flowing out through the $+x$-direction of the box, seen in green in Fig. \ref{fig:lvar}, is used. 

The radial variation of the  CO$^5$BOLD mean BC is derived using mass shells (as described in Sect.~\ref{sect:radvel}) with a suitable mean radial distance. 
An example of a five-year interval of such a mass shell can be seen in the bottom panel of Fig.~\ref{fig:lvar}, in orange. 
For the CO$^5$BOLD x-dir BC however we cannot derive the radial variation in a similar way. 
Instead the radial gas dynamics at $1.5R_\star$(seen in the lower panel to the left in Fig. \ref{fig:rvcomp}) are imitated, by constructing a radial piston that leads to similar shockwave velocities and shock propagation timing in the DARWIN atmosphere. 
This is possible as the gas dynamics in this region for the DARWIN models, before any dust condenses, mainly depend on the inner boundary conditions. 
The amplitude and shape of the radial inner boundary condition can therefore be tuned, so the shock velocities and shock timing in the CO$^5$BOLD atmosphere is reconstructed in the DARWIN model. 

An example of this can be seen in Fig. \ref{fig:recon}. 
The upper panel shows the radial velocity at $1.5R_\star$ for the CO$^5$BOLD model \textit{st28gm06n26}. 
The middle panel shows the boundary condition constructed to imitate this radial-velocity profile, which is then used in a DARWIN model (\textit{l70t27}) with the same stellar parameters as the \textit{st28gm06n26} model. 
The lower panel of Fig. \ref{fig:recon} shows the resulting radial velocity of the DARWIN model \textit{l70t27} at $1.5R_\star$, using the inferred boundary condition, in dashed green. 
The original radial velocity in one direction of the CO$^5$BOLD model \textit{st28gm06n26} is also shown in grey. 

Using a boundary condition inferred from the 3D model, the reconstructed radial velocity in the DARWIN model emulates the radial velocity from the CO$^5$BOLD model well. 
The timing of the shockwaves passing $R = 1.5R_\star$ is similar in both models, however the amplitudes can differ in some instances. 
This scheme of reconstructing the CO$^5$BOLD behaviour seems to capture the overall behaviour of the gas dynamics in CO$^5$BOLD in one direction. 
Radial inner boundary conditions for the DARWIN models are extracted in the same way for all the CO$^5$BOLD models (listed in Table~\ref{tab1}).

\subsection{Comparison of amplitudes}

   \begin{figure}
   \centering
   \includegraphics[width=\hsize]{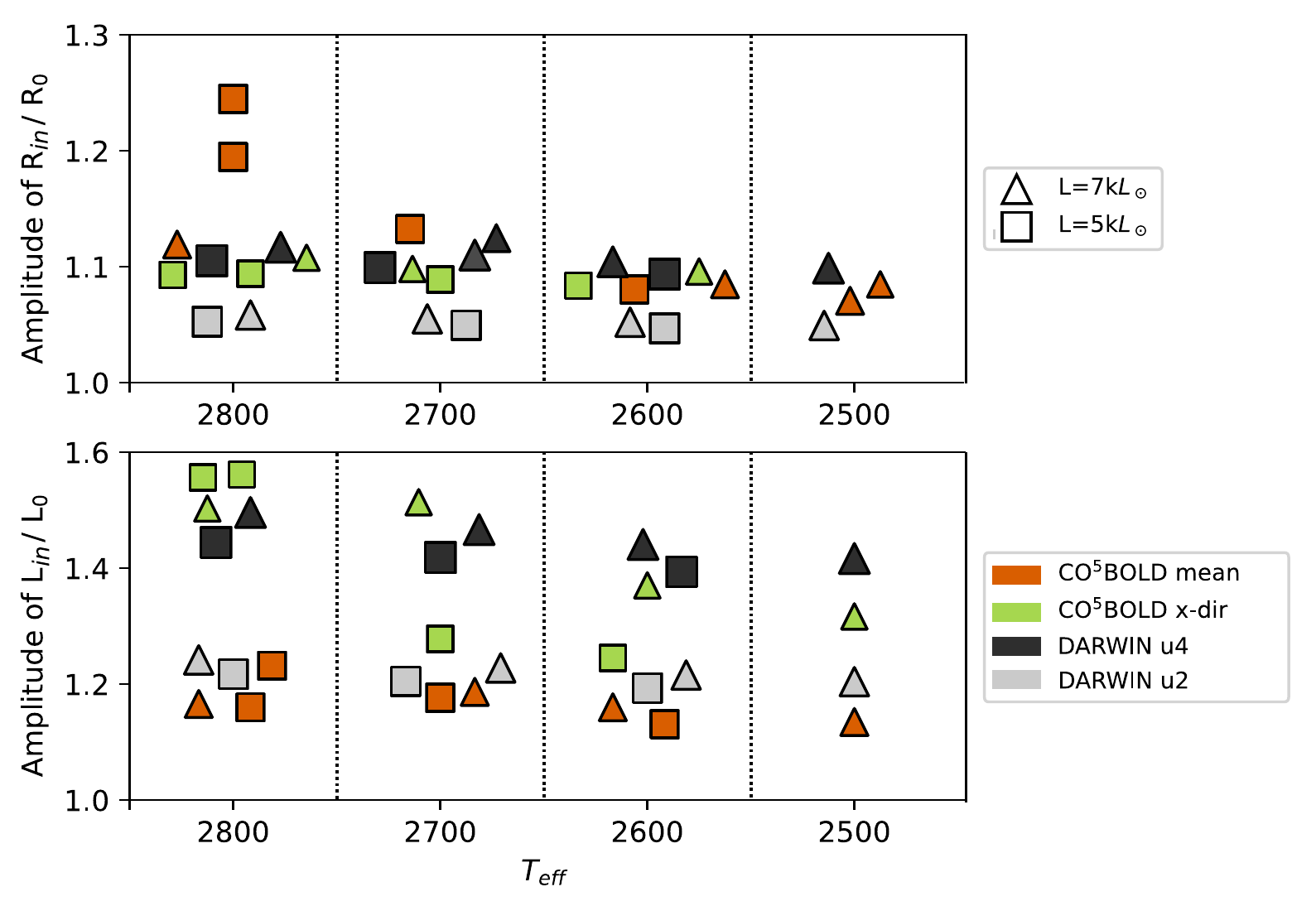}
      \caption{Comparison of the amplitudes of the radial variation and the luminosity variation, for the boundary conditions used in the different DARWIN models. Each of the boxes separated by dashed lines contain models with the same effective temperature, with a small random shift along the $x$-axis for better separation.}
         \label{fig:ramp}
   \end{figure}

The amplitudes of the Fourier series derived in the previous section can be compared with amplitudes used in the standard DARWIN models, set by the free parameters $\Delta u_p$ and $f_L$. 
As mentioned in Sect. \ref{sect:ib}, $\Delta u_p$ is a measure of the maximum velocity at the inner boundary and $f_L$ is a factor which scales the variation of the luminosity. 

The DARWIN u2 and u4 BC amplitudes can be compared to those derived from the CO$^5$BOLD models, where these amplitudes are the result of the pulsations. 
In Fig. \ref{fig:ramp} the derived values for these parameters are compared to the standard ones used in the DARWIN models. 
The amplitudes shown for the CO$^5$BOLD BC are the maximum amplitudes achieved during the full simulated time span.

As can be seen in Fig. \ref{fig:ramp}, showing amplitudes of $ R_\mathrm{in} / R_0$ and $L_\mathrm{in} / L_0$, amplitudes assumed in the DARWIN models are similar to those derived from CO$^5$BOLD models. 
For the amplitude of the radial variation seen in the upper panel in Fig. \ref{fig:ramp}, the DARWIN u2 models result in a variation of around $5\%$ while DARWIN u4 models result in a variation of $10\%$.
The $R_\mathrm{in}/R_0$ amplitudes of the CO$^5$BOLD BC are in this range, both for the derived mean models and for the reconstructed inner boundary conditions of the one-direction models, with the exception of the mean models with $T_\mathrm{eff} = 2800$K and $L=5000L_\odot$. 
These two models reach a maximum amplitude of almost $20\%$.

For the $L_\mathrm{in} / L_0$ amplitude, the DARWIN models vary by $20\%$ for the u2 BC and $40\%$ for the u4 BC.  
The luminosity amplitude of the CO$^5$BOLD BCs varies greatly, by $50\%$ for the most extreme models down to less than $20\%$. 
Overall the range of values for the BCs derived from the CO$^5$BOLD models are in a similar range to those assumed for the standard DARWIN u2 and u4 models.

\subsection{Results - Wind properties}

   \begin{figure}
   \centering
   \includegraphics[width=\hsize]{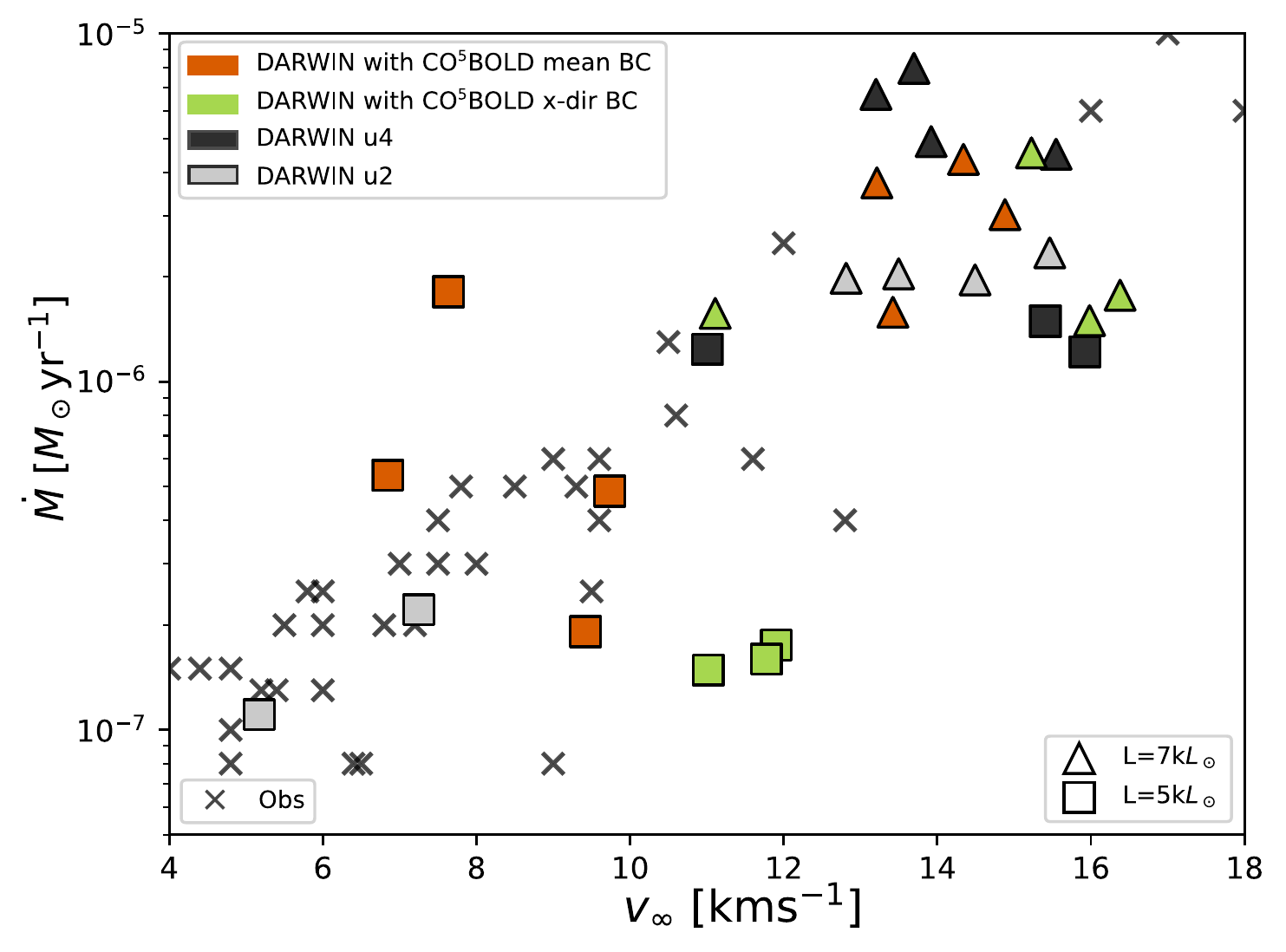}
      \caption{The velocities and mass-loss rates for the different DARWIN models, using both standard and CO$^5$BOLD-derived boundary conditions (squares and triangles). The crosses are mass-loss rate and wind velocity observations by \cite{olofsson_mass_2002} and \cite{gonzalez_delgado_``thermal_2003}.}
         \label{fig:veldmdt}
   \end{figure}

   \begin{figure}
   \centering
   \includegraphics[width=\hsize]{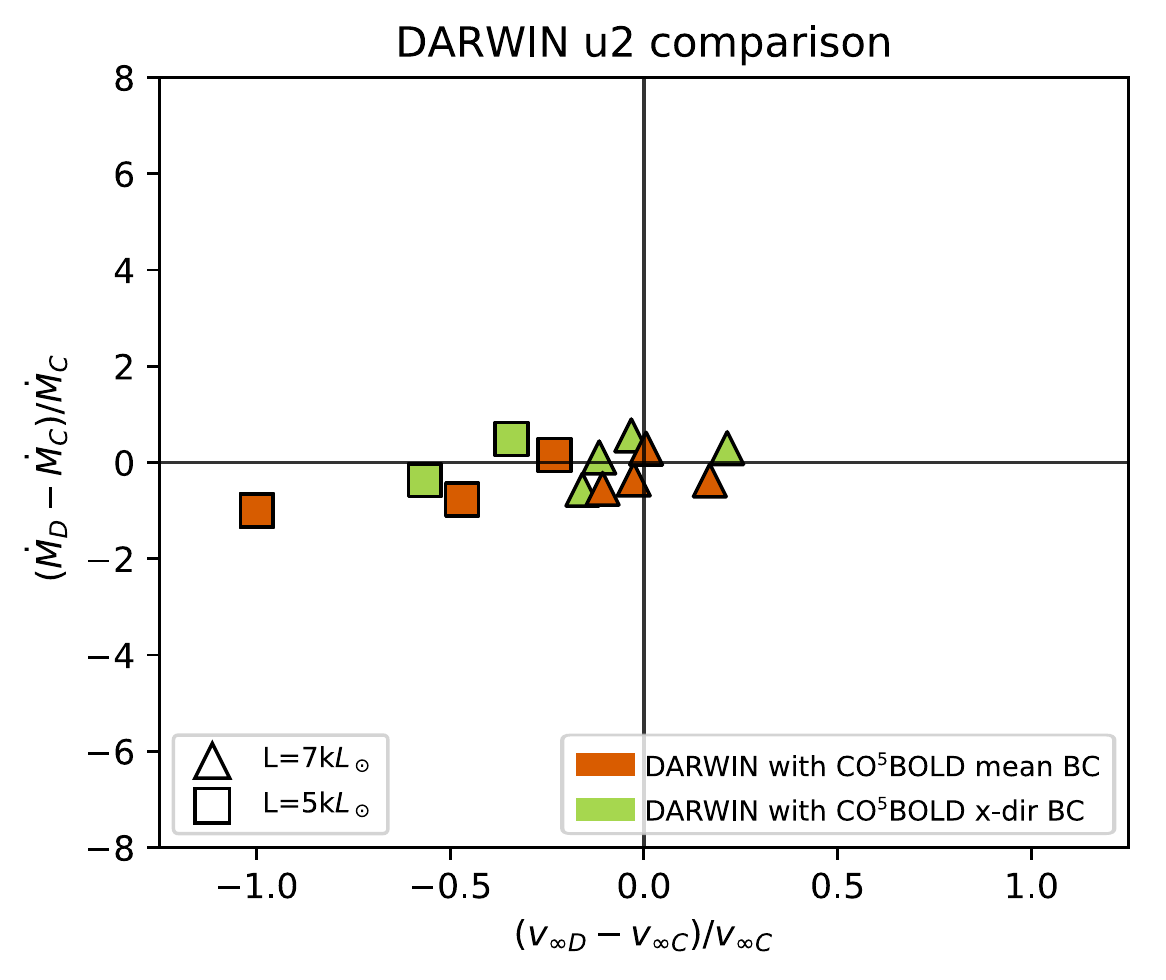}
   \includegraphics[width=\hsize]{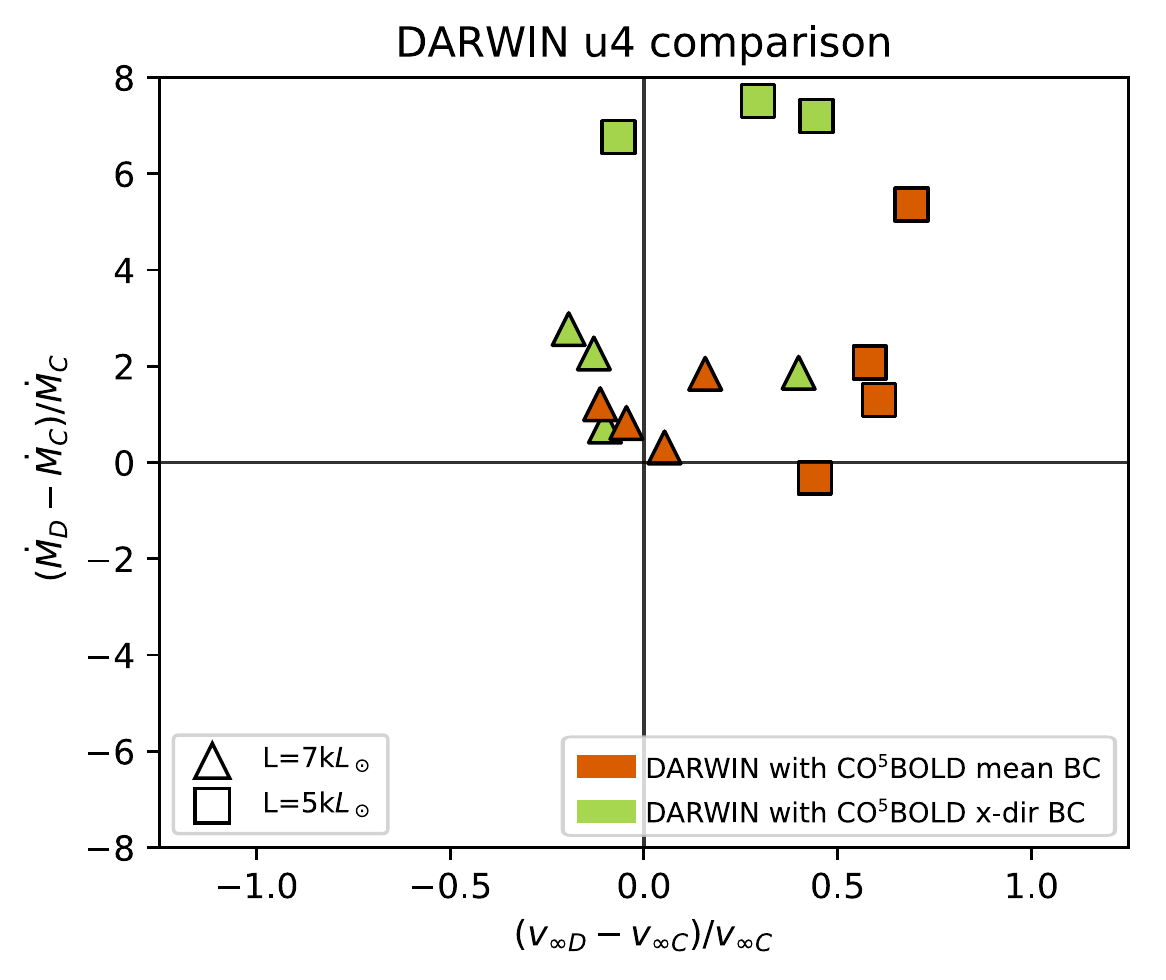}
      \caption{\textit{Top -} The difference in mass-loss rates and wind velocities between the DARWIN models with CO$^5$BOLD boundary conditions and the standard DARWIN models, for u2.
      \textit{Bottom -} The relative differences in mass-loss rates and wind velocities between the DARWIN models with CO$^5$BOLD boundary conditions and the standard DARWIN models, for u4.
      }
         \label{fig:diffcomp1}
   \end{figure}

The wind properties of the standard DARWIN u2 and u4 models are compared to DARWIN models with inner boundary conditions derived from the CO$^5$BOLD models. 

The parameter combination $L=5000L_\star$ and $T_\mathrm{eff}=2800$K is just on the verge of where standard DARWIN models develop a wind. 
The DARWIN u2 BC for this temperature and luminosity does not have a wind at all, and the model with  CO$^5$BOLD x-dir BC only produces an intermittent outflow. 
Further, the CO$^5$BOLD model \textit{st28gm07n001}, with $L=10$ $028L_\star$ and $T_\mathrm{eff}=2506$K, is highly irregular and extended, with a radius close to the size of the computational box.
As discussed in \cite{freytag_global_2017}, this model would most likely need to be simulated with a larger computational box and is therefore here deemed as unreliable. 
These three models are disregarded in the following analysis. 

\subsubsection{Averaged mass-loss rates and wind velocities}

Figure \ref{fig:veldmdt} shows the time-averaged mass-loss rates and wind velocities of the models.
The crosses are observational values of mass-loss rates and wind velocities, from \cite{olofsson_mass_2002} and \cite{gonzalez_delgado_``thermal_2003}.
There is a clear separation between the models with $7000L_\star$ and those with $5000L_\star$.
Increasing the luminosity generally leads to both higher mass-loss rates and higher velocities.
Overall the models seem to fit the trend predicted by the observations (black crosses). 
There are however differences between the different schemes. 
The standard DARWIN u4 models have the highest mass-loss rates and wind velocities. 
The two different sets of DARWIN models using CO$^5$BOLD BC produce dynamical properties closer to that of the DARWIN u2 models.

For a better comparison of the results, the differences in mass-loss rates and wind velocities are plotted in Fig. \ref{fig:diffcomp1}.
The different plots for the DARWIN u2 models and the DARWIN CO$^5$BOLD BC models are seen in the top panel of Fig. \ref{fig:diffcomp1}, with DARWIN u4 models and the DARWIN CO$^5$BOLD BC models in the bottom panel of Fig. \ref{fig:diffcomp1}. 

The mass-loss rates of the DARWIN u2 models correspond closely to the mass-loss rates of CO$^5$BOLD BC models, which agree mostly within a factor of two.
While the wind velocities agree well for the CO$^5$BOLD BC models with $7000L_\star$, there is a larger spread for those with $5000L_\star$.
This is likely because the combination of parameters for models with $5000L_\star$ are close to where the DARWIN models no longer develop a wind.
Any changes to the boundary condition of these models have large effects on the amount of dust formed, and how effective the wind is, which is the most likely explanation for the deviating wind velocities.

From the bottom panel of  Fig. \ref{fig:diffcomp1} its clear that the u4 models have consistently higher mass-loss rates, when compared to the models using CO$^5$BOLD BC.
For the  $5000L_\star$ models, the difference in mass-loss rates is up to almost a factor of eight.
The DARWIN models with CO$^5$BOLD mean BC have mass-loss rates closer to that of the DARWIN u4 models, so the difference is largest when comparing to the DARWIN models with CO$^5$BOLD x-dir BC.
The wind velocity varies less, between $-20\%$ and $+60\%$.

Overall, the the mass-loss rates of the DARWIN u2 models agree better with the DARWIN CO$^5$BOLD BC models, for these combinations of stellar parameters.

\subsubsection{Time evolution of the wind}
\label{sect:te}

   \begin{figure}
   \centering
   \includegraphics[width=\hsize]{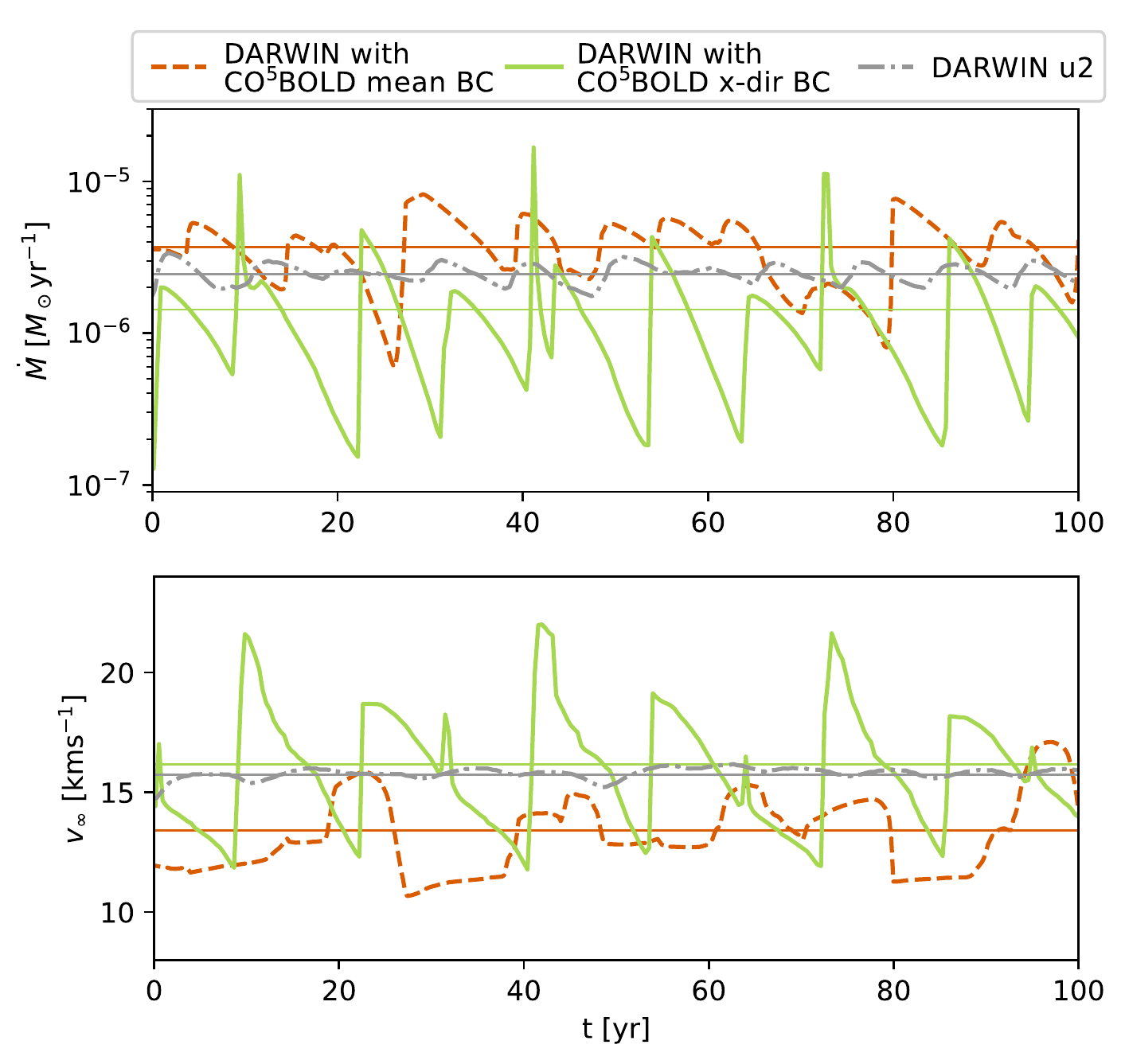}
      \caption{The variation of the mass-loss rate and the velocity at $25R_\star$ for the DARWIN \textit{l70t27} models with CO$^5$BOLD BC compared to the standard DARWIN u2 BC (the standard DARWIN u4 model shows similar steady wind). The horizontal lines indicate the mean values for each model, in mass-loss rates and wind velocities. }
         \label{fig:timeevo}
   \end{figure}

While the average mass-loss rates and wind velocities agree quite well between the standard DARWIN u2 models and the models with CO$^5$BOLD BC, the time-varying behaviour of the lower atmosphere of the 3D CO$^5$BOLD models is reflected in the resulting wind.
Figure \ref{fig:timeevo} shows the temporal variation of the wind velocity and mass-loss rates at the outer boundary (at $25R_\star$), for the \textit{l70t27} model using three different boundary conditions.
The \textit{l70t27} model (the model also examined in Sect. \ref{sect:comp}) is here used as an example for the general behaviour shown for all different stellar parameters, when using different inner boundary conditions.

The standard DARWIN u2 model, with regular inner boundary, results in an almost steady wind where the temporal variations in both mass-loss rate and wind velocity are very small (see the grey line in Fig. \ref{fig:timeevo}). 
The DARWIN u4 model (not shown here) results in a similar steady wind with little temporal variation.
This is in stark contrast with the DARWIN models with the two CO$^5$BOLD BCs, where a cycle-to-cycle variation occurs. 
The temporal variation in radial amplitudes leads to more or less material being levitated to distances where the gas can condense into dust in each cycle, which influences the efficiency of the wind driving. 
Similarly, the temporal variation in luminosity amplitudes leads to different radiative pressure on the dust, which again has importance for the resulting wind. 

The resulting winds of the DARWIN models with CO$^5$BOLD BC therefore show a variation in both the wind velocity and the mass-loss rate, on timescales of 10-20 years, at the outer boundary set to $25R_\star$.
Similar variable wind properties can be present in the standard DARWIN models as well, especially for larger values of $\Delta u_p$ for the C-star models \citep[see][]{eriksson_synthetic_2014}.
Such models have larger ballistic timescales, which can lead to more complex behaviour in the atmosphere with, for example, dust only forming every other cycle. 
For the sample of models examined in this paper however, the DARWIN models with CO$^5$BOLD BC have consistently significantly larger time variations in both mass-loss rates and wind velocities than the corresponding standard DARWIN models.

\section{Discussion}

When comparing the dust-free lower atmosphere of the 3D CO$^5$BOLD models with the spherically symmetric DARWIN models we find some significant differences. 
The DARWIN standard models use a sinusoidal boundary condition for the radial and luminosity variation, resulting in a regular behaviour at the inner boundary and in the lower atmosphere. 

Results from the CO$^5$BOLD models however indicate that this assumption is a simplification.
While the shockwaves in the 3D simulations are of a global scale, they are not spherically symmetric.
Each shockwave contains a distribution of velocities, so the radial gas velocities will depend on the direction. 
Similarly the luminosity averaged over all the sides of the computational box  and radial variation averaged over spherical shells of the CO$^5$BOLD models varies from cycle to cycle.

When comparing the amplitudes of variations in the CO$^5$BOLD models with those used in the standard DARWIN models, the CO$^5$BOLD models typically have a $R_\mathrm{in} / R_0$ amplitude similar to the DARWIN u4 models, but a $L_\mathrm{in}/L_0$ amplitude similar to the  DARWIN u2 models.
This is the general trend, however, and some CO$^5$BOLD models do reach the higher luminosity amplitudes. 
When comparing observed $\Delta$K magnitudes with synthetic $\Delta$K magnitudes, using DARWIN C-type AGB star models, it has previously been shown that a luminosity amplitude parameter of $f_l = 2$, as used here, results in a reasonable fit \citep{eriksson_synthetic_2014}.
 No systematic comparisons have yet been made for DARWIN M-type models.

When using the CO$^5$BOLD BC in the DARWIN models, the DARWIN u2 models agree quite well with the average mass-loss rates and the wind velocities. 
However, the non-sinusoidal  behaviour of the lower boundary of the DARWIN models with CO$^5$BOLD input affects the resulting dynamics. 
When the irregular radial and luminosity variations interact with other physical processes with different time-scales, such as dust formation,  wind acceleration and gravity, the resulting mass-loss rates and wind velocities become more complex than the standard case. 
In the example in Sect. \ref{sect:te} there is an order-of-magnitude difference in the mass-loss rates and the wind velocity change of 10km s$^{-1}$, on timescales of 10-20 years.
Subsequently, when faster-moving material collides with slower gas, higher-density pile-ups and small-scale structures are created in the stellar wind.
This is a consequence of the time-varying behaviour predicted in the CO$^5$BOLD models and emulated in the DARWIN models with CO$^5$BOLD boundary conditions. 

Shell structures in the circumstellar envelope have also been shown to occur due to drift between the dust and the gas, which at least for the case of C-stars decouples in the envelope \citep[see][]{simis_origin_2001}.
Such arc-like structures have been observed in IRC+10216 \citep{mauron_multiple_1999},  for example, with timescales corresponding to 200-800 years.
The dust grain sizes in models of this phenomenon are however assumed to be several orders of magnitude smaller than the silicate grains in M-stars, which highly influence the distances at which this decoupling occurs.
It is unclear if a similar decoupling occurs for M-type AGB stars, and what the influence of a variable wind would be.

The timescales of the variations seen in the winds of the DARWIN models with CO$^5$BOLD BC are of the order of 10-20 years. 
This is about an order of magnitude larger than the period ($\sim$1 year) but an order of magnitude smaller than the observed structures.
It is unclear if the model results can be directly compared to the observations however, as the models only simulate the wind out to 25 R$_\star$.
Further investigation into this would be needed to make any decisive
conclusions, with wind-wind interaction models that reach further out into the circumstellar envelope.

Another consequence of the irregular shockwaves is that the amount of dust that condenses depends on both direction and cycle. 
This has previously been investigated for 3D AGB star models, by \cite{freytag_three-dimensional_2008}, which included passive dust formation of amorphous carbon grains without considering the coupling with the radiation field in their calculations. 
They found that dust forms in non-spherical structures around the star, however since the dust was not radiatively accelerated no wind was induced in these models.
It is here found that the shockwaves typically only cover around $70\%$ of the stellar surface at $R=1.5R_\star$, indicating again that the distribution of newly formed dust in the close vicinity of the AGB star should not be spherically symmetric but only cover this amount or less of the stellar surface. 
Such non-spherical dust structures are also observed by \cite{ohnaka_clumpy_2016,ohnaka_clumpy_2017}, for example.
To fully understand the implications of anisotropic  dust formation on the resulting wind, however, 3D models with full coupling between gas and dust are needed.

The standard DARWIN models do not take the distribution of the radial shock velocities, as seen in CO$^5$BOLD models, into account. 
It is therefore quite possible that stellar parameter combinations that do not result in a dust-driven wind when modelled with the standard DARWIN boundary condition scheme might in reality actually produce such a wind.
While the mean of the radial shock velocity distribution in such a star, which is in essence what the DARWIN code tries to emulate, is too low, the higher-end tail velocities might levitate material far enough into the inner atmosphere. 
Dust might then condense due to this higher-velocity material, creating a possibly intermittent dust-driven outflow. 
There are currently no 3D models with suitable parameter combinations to investigate this. 
The grid of CO$^5$BOLD 3D models should therefore be expanded to include stellar parameter combinations that do not produce a dust-driven wind in the standard DARWIN models.
Such models may cast light on whether or not a dust-driven wind can be triggered for less-evolved AGB stars.

\section{Summary and conclusion}

We try to estimate the effects of a non-spherical star, as predicted by CO$^5$BOLD 3D interior models, on the wind properties of AGB stars.
To summerize the result:

\begin{itemize}
\item The gas velocities in a shock front in CO$^5$BOLD models are not uniform, but  rather a distribution of velocities (Fig. \ref{fig:veld}). This might be important for the wind-driving in less-evolved AGB stars and could potentially lead to a weak dust-driven wind earlier on the AGB than predicted by the DARWIN models. 
\item Only about $70\%$ of the full surface of the CO$^5$BOLD models is covered by shockwaves during a cycle (Fig. \ref{fig:area}). This varies from cycle to cycle. 
\item The CO$^5$BOLD models do however show sporadic variations both in space and in time for the gas velocities and luminosity amplitudes, in contrast to the spherically symmetric DARWIN models where these quantities are assumed to vary sinusoidally (Fig. \ref{fig:veld}, \ref{fig:lvar} and \ref{fig:rvcomp}). 
\item The amplitudes of the luminosities and the gas velocities in the close, dust-free atmosphere of the CO$^5$BOLD 3D models are similar to those assumed for the DARWIN dynamical atmosphere models (Fig. \ref{fig:ramp}).
\item When using the CO$^5$BOLD interior models as input for the DARWIN wind models the resulting average dynamical properties agreed well with the standard DARWIN u2 models (Fig. \ref{fig:veldmdt} and top panel of \ref{fig:diffcomp1}). 
The DARWIN u4 models have consistently higher mass-loss rates when compared to the DARWIN models using CO$^5$BOLD BC.
\item DARWIN models with CO$^5$BOLD input show large variations with time, and mass-loss rates could vary by an order of magnitude and the wind velocity  with $50\%$ over a timescale of 10-20 years (Fig. \ref{fig:timeevo}). Such large variations in the density  of the wind might cause observable small-scale structures in the circumstellar envelope. 
\end{itemize}

While the average dynamical properties were similar for the standard DARWIN models and the DARWIN models with CO$^5$BOLD input, the anisotropic star predicted by the CO$^5$BOLD models affect the wind.
To see if such variability of the wind properties results in observable structures, more investigation is needed. 

The non-uniform behaviour of the shockwaves shown by the CO$^5$BOLD models might induce a dust-driven wind in less evolved AGB stars than predicted by the DARWIN models. 
CO$^5$BOLD models with relevant stellar parameters in combination with DARWIN models are needed to further explore this.

Ideally, 3D models that include dust-driven winds should be used for studying the mass loss of AGB stars. 
However, such models are not available yet.
Additionally, the computational time of such models would be several orders of magnitude higher than for the current spherical models, making it impractical, for example, in extensive wind model grids used to derive mass-loss rates, which span a wide range of stellar parameters, as necessary for stellar-evolution modelling. 
The methods developed in this paper, deriving pulsation properties from the CO$^5$BOLD models, can be used to mimic the effects of the pulsations and giant convection cells and avoid free parameters in DARWIN models.
If the current grid of CO$^5$BOLD 3D models were expanded, it would therefore be possible to infer inner boundary conditions from a sparse grid of 3D models onto a well-sampled grid of DARWIN models.

\begin{acknowledgements}
This work has been supported by the Swedish Research Council (Vetenskapsr{\aa}det) and by a stipend provided by the Sch{\"o}nberg donation. The computations of the models were performed on resources provided by the Swedish National Infrastructure for Computing (SNIC) at UPPMAX.
\end{acknowledgements}

\bibliographystyle{aa} 
\bibliography{ref,ref2}

\end{document}